\documentclass[twocolumn]{aastex631}
\usepackage{graphicx}
\usepackage{txfonts}

\shorttitle{The repulsive effect in close pairs}

\newcommand{\vecs}{\mbox{\boldmath $s$} {}}
\newcommand{\vecF}{\mbox{\boldmath $F$} {}}
\newcommand{\vecr}{\mbox{\boldmath $r$} {}}

\newcommand{\vecv}{\mbox{\boldmath $v$} {}}

\newcommand{\vecG}{\mbox{\boldmath $\Gamma$} {}}
\usepackage{bm}
\usepackage{multirow}
\usepackage{xcolor}
\usepackage{amssymb}	

\definecolor{sonia}{HTML}{b34db3}
\definecolor{javier}{HTML}{4db380}
\definecolor{fred}{HTML}{3cc1f2}

\shorttitle{The repulsive effect}
\shortauthors{S\'anchez-Salcedo, Masset \& Cornejo}

\begin{document}

\title{A close pair of orbiters embedded in a gaseous disk: the repulsive effect}

\correspondingauthor{F. J. S\'anchez-Salcedo}
\email{jsanchez@astro.unam.mx}

\author[0000-0003-2416-2525]{F. J. Sánchez-Salcedo}
\affiliation{Instituto de Astronom\'{\i}a, Universidad Nacional Aut\'{o}noma de M\'{e}xico, AP 70-264, 
Mexico City 04510, Mexico}
\author[0000-0002-9626-2210]{F. S. Masset}
\affiliation{Instituto de Ciencias F\'{\i}sicas, Universidad Nacional Aut\'onoma de M\'exico, Cuernavaca, Morelos, Mexico}

\author[0009-0000-7637-6694]{S. Cornejo}
\affiliation{Instituto de Ciencias F\'{\i}sicas, Universidad Nacional Aut\'onoma de M\'exico, Cuernavaca,
Morelos, Mexico and Max Planck Institute for Solar System Research, Justus-von-Liebig-Weg 3, D-37077 G\"ottingen, Germany}

\begin{abstract}
We develop a theoretical framework and use two-dimensional hydrodynamical simulations to study the repulsive effect between two close orbiters embedded in an accretion disk. We consider orbiters on fixed Keplerian orbits with masses low enough to open shallow gaps. The simulations indicate that the repulsion is larger for more massive orbiters and decreases with the orbital separation and the disk's viscosity.
We use two different assumptions to derive theoretical scaling relations for the repulsion. A first scenario assumes that each orbiter absorbs the angular momentum deposited in its horseshoe region by the companion's wake. A second scenario assumes that the corotation torques of the orbiters are modified because
the companion changes the underlying radial gradient of the disk surface density.
We find a substantial difference between the predictions of these two scenarios. The first one fails to reproduce the scaling of the repulsion with the disk viscosity and generally overestimates the strength of the repulsion. The second scenario, however, gives results that are broadly consistent with those obtained in the simulations.

\end{abstract}


\section{Introduction}

A massive body embedded in an accretion disk can migrate radially due to the disk torques. For instance, in the accretion disks in the center of active galactic nuclei (AGNs), disk torques may lead to the inward migration of stellar-mass ($\sim 10M_{\odot}$) and intermediate-mass (defined as those with masses between $60$ and $10^{5}M_{\odot}$) black holes (BHs) 
\citep[e.g.,][]{koc11,mck11}.
It has been suggested that $10M_{\odot}$ BHs embedded in the accretion disks of AGNs can accumulate, scatter, and merge in migration traps 
\citep{bel16,sec19,yan19b,mck20}.

In protoplanetary disks, embryos and protoplanets can also migrate inwards or outwards \citep[e.g.,][]{bar13rev,nel18}.
During their migration, planets can be trapped in mean-motion resonance (MMR) 
\citep[e.g.,][]{izi17}.
Nevertheless, observations indicate that planetary pairs in compact multi-planet systems are generally not found in MMR. There is, however, a small population of pairs that are in near-resonance but with a tendency to have orbital period ratios larger
than required for exact first-order MMR \citep{lis11,fab14}. 
Understanding the processes that move systems just out of MMR has been the subject of
numerous works \citep[e.g.][and references therein]{cha22}.

\citet{bar13} suggest that the interactions between a planet and the wake of its companion can cause a ``repulsion'' of the orbits, which could account for the observed shifts
from the nominal commensurated period ratios.
In this scenario, the disk mediates an exchange of angular momentum between the 
planets.
Unlike resonant repulsion effects \citep[e.g.,][]{lit12,cho20},
the interaction with the wake of the companion can induce orbital repulsion 
between the planets without involving the direct gravitational coupling between
them. 
The orbital repulsion due to the ``wake-planet'' interaction appears more effective 
when the planets open partial gaps in the disk. 
Interestingly, the wake-planet repulsion between
an inner Jovian planet and an outer super-Earth can halt convergent migration even before they are captured in first-order MMR \citep{pod12}. 
\citet{cui21} study the migration of two super-Earths (planet-to-star mass ratios 
$\sim 10^{-5}$) 
in the inner parts ($\sim 1$ au) of protoplanetary disks \citep[see also][]{ata21}. They include a central cavity 
in the disk surface density, producing a migration trap for the inner planet.
The migration changes from convergent to divergent in the models explored by \citet{cui21}.

The repulsive effect between pairs is relevant not only in interpreting the architectures of planetary systems but it could also help understand the role of the gas in the evolution of a closely-packed pair of BHs in the accretion disks of AGNs and to quantify the chance for the formation of a bound BH binary system \citep{row23}. 

It has been suggested that the repulsion effect seen in the simulations is a consequence of the modification of the corotation torques acting on each orbiter because a fraction of the angular momentum carried by the wake excited by the inner orbiter can be deposited in the coorbital region of the outer orbiter and vice versa, altering the corotation torques \citep{bar13,cui21}.
It seems that this interpretation can explain {\it qualitatively} the repulsion effect, but a quantitative analysis of this hypothesis would be useful for full satisfaction.

In fact, \citet{kan20} study the migration of
two gap-opening planets and show that the transition from convergent to divergent migration can be accounted for without invoking any repulsion mechanism.
In the first stage, the planets undergo convergent migration because they have insufficient time to clear their gaps. After being captured in resonance, they open their gaps, and the migration can become divergent.
In this scenario, the migration will switch from convergent to divergent if the migration rate of the inner planet in a steady state, calculated as if it were at isolation in the disk, is larger than that of the outer planet.

To gain a deeper insight into the nature and magnitude of the disk-mediated repulsion effect, in Section \ref{sec:theory}, we  estimate the
strength of the orbital repulsion adopting two different hypothesis. 
In Section \ref{sec:simulations}, we compare the predicted scalings with those obtained
from two-dimensional simulations of a pair of orbiters to determine which hypothesis
is more appropriate. A discussion and summary of the results are given in Sections \ref{sec:discussion} and \ref{sec:summary}, respectively.

\section{The repulsion effect: theoretical aspects}
\label{sec:theory}
We consider a pair of massive bodies embedded in the midplane of an accretion disk around a central object with mass $M_{\bullet}$. 
These orbiters can be either two planets in a protoplanetary disk or a pair of BHs in the 
AGN accretion disk. 
The orbiters have masses $M_{1}$ and $M_{2}$, semi-major axes $a_{1}$ and $a_{2}$ 
 and eccentricities $e_{1}$ and $e_{2}$, respectively.
We will assume that $a_{1}<a_{2}$ so that orbiter $1$ is the inner body.
In this paper, our aim is to shed light on the disk-orbiters interaction taking into
account that the disk is disturbed by the companion. For this purpose, 
we will ignore the gravitational forces between the orbiters and consider only the disk 
force on each perturber arising due to the density perturbations induced in the disk by 
both bodies.

Each body excites density waves that carry angular momentum. These
waves can modify the structure of the disk.
Due to the wake of the companion, the torque acting on orbiter $j$, denoted by $T_{j}$, 
will change by an amount $\delta T_{j}$. Throughout this paper, we will use the 
convention that $T_{j}$ is positive (negative) if the body gains (loses) angular momentum.

It is expected that a fraction $\lambda_{1}$ of the one-sided torque excited in the 
disk by the orbiter $2$, denoted by $T_{1s,2}$, is absorbed by orbiter $1$ and vice versa. Hence, $\delta T_{1}=-\lambda_{1}|T_{1s,2}|$,
whereas $T_{2}$ will change by an amount $\delta T_{2}=\lambda_{2}|T_{1s,1}|$.

Two different cases with different assumptions are used to derive
$\lambda_{1}$ and $\lambda_{2}$. 
In the first scenario, we assume that
each orbiter absorbs the angular momentum flux, excited by its companion, that is
deposited in its horseshoe region \citep[e.g.,][]{bar13,cui21}. 
For brevity, we will refer to this case as ``angular-momentum modified torques''
(hereafter AMMT). 
In Appendix \ref{sec:app_AMMT}, we provide the formulae to evaluate $\lambda_{1}$ and 
$\lambda_{2}$ in AMMT. 

The second scenario assumes that the corotation torque on each orbiter is modified because the companion changes the radial profile of the disk (or, more specifically, the vortensity gradient). In this scenario, the outer orbiter will feel a larger (positive) corotation torque
than when it is at isolation in the disk, 
because it lies at the outer edge of the gap opened by the inner orbiter. On the contrary,
the corotation torque on the inner orbiter decreases because it lies on the inner edge of the gap opened by the outer orbiter.
We will refer to this case as ``density-gradient modified torques'' (hereafter DGMT).
In Appendix \ref{sec:app_DGMT}, we give $\lambda_{1}$ and $\lambda_{2}$  in DGMT.
In Section \ref{sec:predictions}, we will see that AMMT and DGMT provide different predictions.

For simplicity, in the derivation of $\lambda_{1}$ and $\lambda_{2}$,  we will assume 
that the orbital eccentricity of both orbiters remains small. This is generally seen in the simulations. For instance, the orbital eccentricity of the planets reaches values up to $e\sim 0.01$
in the simulations of \citet{cui21}, and up to $e\sim 0.03$ in the simulations of \citet{bar13}.

We warn that the theoretical estimates for $\lambda_{1}$ and $\lambda_{2}$ 
in the Appendices \ref{sec:app_AMMT} and \ref{sec:app_DGMT} are accurate
for perturbers that, when they are alone in the disk, open a partial gap, that is, 
$\Sigma_{{\rm gap},j}/\Sigma_{{\rm un},j}\geq 0.65$.
Here $\Sigma_{{\rm gap},j}$ is the disk surface density at the bottom of the gap created by orbiter $j$, and $\Sigma_{{\rm un},j}$ is the unperturbed disk surface density at $R=a_{j}$. Using the empirical formula of \citet{duf15} for the gap depth created by a single perturber, we find that this condition implies $q_{j}^{2}\lesssim 10\alpha h^{5}$, where $q_{j}\equiv M_{j}/M_{\bullet}$, $h$ is the disk aspect ratio and $\alpha$ the Shakura-Sunyaev viscosity parameter. 

\subsection{Convergent and divergent migration}

The evolution of the orbital parameters of the orbiter $j$ is determined by the power $P_{j}=\vecv_{j}\cdot \vecF_{d,j}$ and the torque $\vecG_{j}=\vecr_{j}\times \vecF_{d,j}$ acting on perturber $j$, where $\vecr_{j}$ and $\vecv_{j}$ are the position and velocity vectors, respectively. In particular, the evolution of the semi-major axis $a_{j}$ is
\begin{equation}
\frac{da_{j}}{dt}= \frac{2 P_{j} }{\omega_{j}^{2} a_{j}M_{j}},
\end{equation}
where $\omega_{j}$ is the orbital frequency of orbiter $j$.

The ratio of orbital radii $\xi\equiv a_{2}/a_{1}$ would evolve according to
\begin{equation}
\frac{1}{\xi} \frac{d\xi}{dt} = \frac{2  \mathcal{D}}
{\omega_{0} (a_{0}^{3} a_{2})^{1/2}},
\label{eq:dlnxi_dt}
\end{equation}
with
\begin{equation}
\mathcal{D}\equiv \frac{P_{2}}{\omega_{2} M_{2}}-\frac{\xi^{1/2} P_{1}}{\omega_{1}M_{1}}.
\label{eq:def_D}
\end{equation}
In Equation (\ref{eq:dlnxi_dt}), we have written $\omega_{j}=\omega_{0} (a_{0}/a_{j})^{3/2}$, where $\omega_{0}$
is the orbital frequency at a reference radius $a_{0}$.
Note that $\mathcal{D}$ has dimensions of specific torque. 
We will say that the migration is convergent if $d\xi/dt<0$, i.e. if $\mathcal{D}<0$,
and divergent if $\mathcal{D}>0$.

We may express $\mathcal{D}$ as the sum of two terms
\begin{equation}
\mathcal{D}=
\mathcal{D}_{0}+\mathcal{D}_{\rm I},
\end{equation} 
where $\mathcal{D}_{0}$ is the value of $\mathcal{D}$ under the assumption that
the torques the orbiters experience are not affected by the presence of their companion,
and $\mathcal{D}_{\rm I}$ represents the change of $\mathcal{D}$ caused by the presence 
of the companion.
We will refer to $\mathcal{D}_{\rm I}$ as the interaction offset.

If the orbiters move on quasi-circular orbits, then $P_{j}/\omega_{j}= T_{j}$ and
\begin{equation}
\mathcal{D}_{\rm I}= \lambda_{2}M_{2}^{-1}|T_{1s,1}|+\lambda_{1} \xi^{1/2}M_{1}^{-1}|T_{1s,2}|.
\label{eq:Dint_basic}
\end{equation}
In both AMMT and DGMT scenarios, it holds that $\lambda_{1}$ and $\lambda_{2}$ are positive and hence $\mathcal{D}_{\rm I}>0$ (repulsive effect).
The one-side torque excited in the disk by orbiter $j$ is given by
\begin{equation}
|T_{1s,j}|=f_{0}\gamma_{j}h^{-3}q_{j}^{2} a_{j}^{4}
\omega_{j}^{2}\Sigma_{\rm un,j},
\label{eq:one_sided}
\end{equation}
where  
\begin{equation}
\gamma_{j}=\left(1+\frac{f_{0}}{3\pi}\frac{q_{j}^{2}}{\alpha h^{5}}\right)^{-1},
\label{eq:gamma_j}
\end{equation}
 with $f_{0}\simeq 0.45$ 
\citep[e.g.,][]{duf15}.

\begin{figure}
  \includegraphics[angle=0,width=81mm,height=139mm]{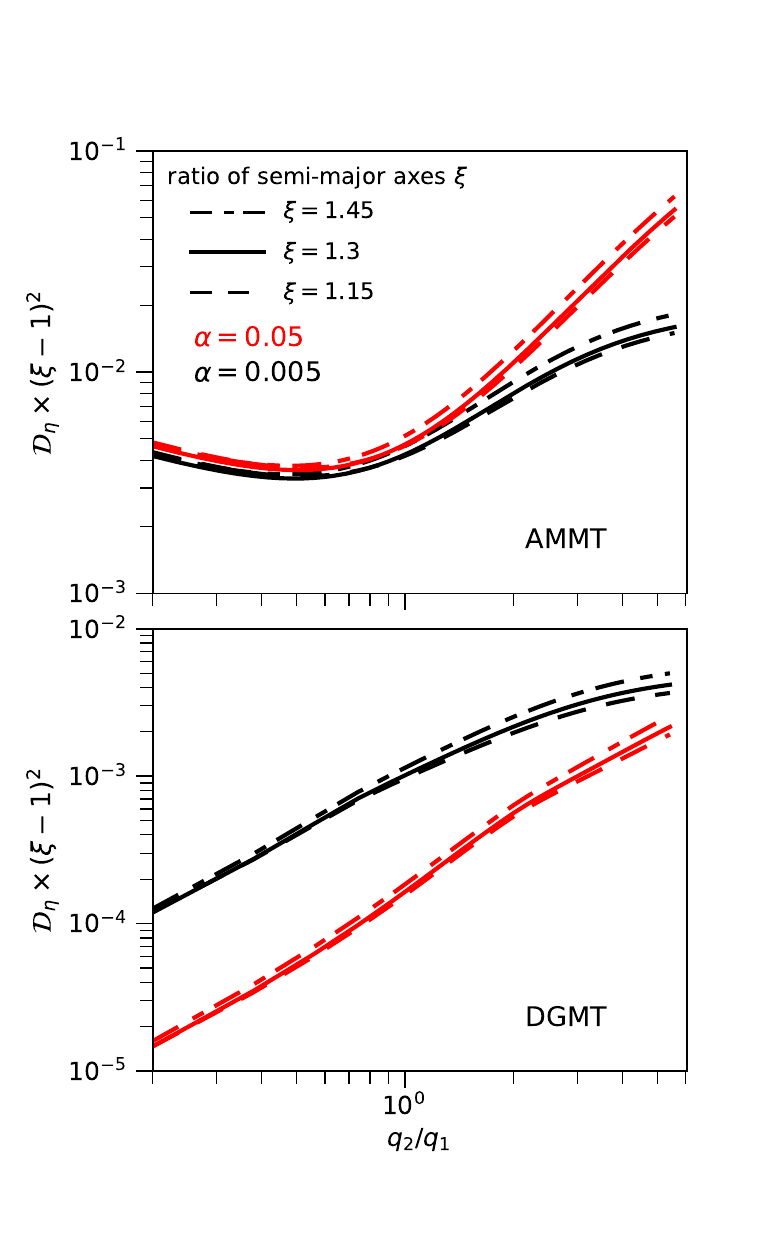}
 \caption{$(\xi-1)^{2}\mathcal{D}_{\eta}$ versus $\eta\equiv q_{2}/q_{1}$ assuming
AMMT (top panel) and DGMT (bottom panel), for
$\alpha=0.005$ (black curves) and for $\alpha=0.05$ (red curves). 
The ratio of semi-major axes $\xi$ varies from curve to curve according to the line style given in the upper-left corner of the
top panel.
For all curves $q_{1}=1.5\times 10^{-5}$ and $h=0.028$.
}
 \label{fig:Dint_eta}
 \end{figure}

\subsection{Theoretical predictions}

\label{sec:predictions}
We first explore how $D_{\rm I}$ depends on the orbiters' separation.
For simplicity, we will assume that $\Sigma_{\rm un}(R)=\Sigma_{0}(a_{0}/R)^{p}$, where $\Sigma_{0}$ is the surface density at the reference radius $a_{0}$.

Combining Eqs. (\ref{eq:Dint_basic}) and (\ref{eq:one_sided}), the interaction offset can be 
written as
\begin{equation}
\mathcal{D}_{\rm I}= f_{0}h^{-3} \omega_{0}^{2} a_{0}^{3+p} a_{1}^{1-p}
 \Sigma_{0} M_{\bullet}^{-1} q_{1}\mathcal{D}_{\eta}(\xi),
\label{eq:Dint_final}
\end{equation}
where 
\begin{equation}
\mathcal{D}_{\eta}(\xi)= \eta^{2}\lambda_{1}\gamma_{2} \xi^{3/2-p}+\eta^{-1} \lambda_{2}\gamma_{1}
\label{eq:Deta}
\end{equation}
and $\eta\equiv q_{2}/q_{1}$. 
We found empirically that the combination $(\xi-1)^{2}\mathcal{D}_{\eta}$, with 
$\mathcal{D}_{\eta}$ given in Equation (\ref{eq:Deta}), varies little to changes in $\xi$,
implying that, in the range of $\xi$ under consideration, $\mathcal{D}_{\eta}$ depends 
on $\xi$ as $\mathcal{D}_{\eta} \sim (\xi-1)^{-2}$, in both AMMT and DGMT.
This is shown in Figure \ref{fig:Dint_eta}, where we compare $\mathcal{D}_{\eta}$ 
for three different values of $\xi$. We took $q_{1}=1.5\times 10^{-5}$, $h=0.028$ and $p=0.5$.
Indeed, throughout this paper, we will take $p=0.5$ unless otherwise stated.

A clear difference between the models is that, for any value of $\eta$, 
$\mathcal{D}_{\eta}$ is significantly smaller in DGMT than in AMMT. Another difference
is that, at fixed $\xi$ and $\alpha$, $\mathcal{D}_{\eta}$ increases monotonically 
with $\eta$ in DGMT, whereas it presents a minimum at $\eta\simeq 0.5$ in AMMT. 

The dependence of $\mathcal{D}_{\eta}$ with $\alpha$ in AMMT also differs from that in DGMT. In AMMT, $\mathcal{D}_{\eta}$ for $\alpha=0.05$ is larger than it is for $\alpha=0.005$, especially for $\eta>2$. The reason is that the angular momentum deposited by the outer perturber decreases as its mass increases because it opens a deeper gap. However, DGMT predicts a reduction of the repulsive effect because radial gradients in the disk surface density induced by the companion become smaller as viscosity increases.

\begin{figure}
\hskip 0.0cm
 \includegraphics[angle=0,width=92mm,height=99mm]{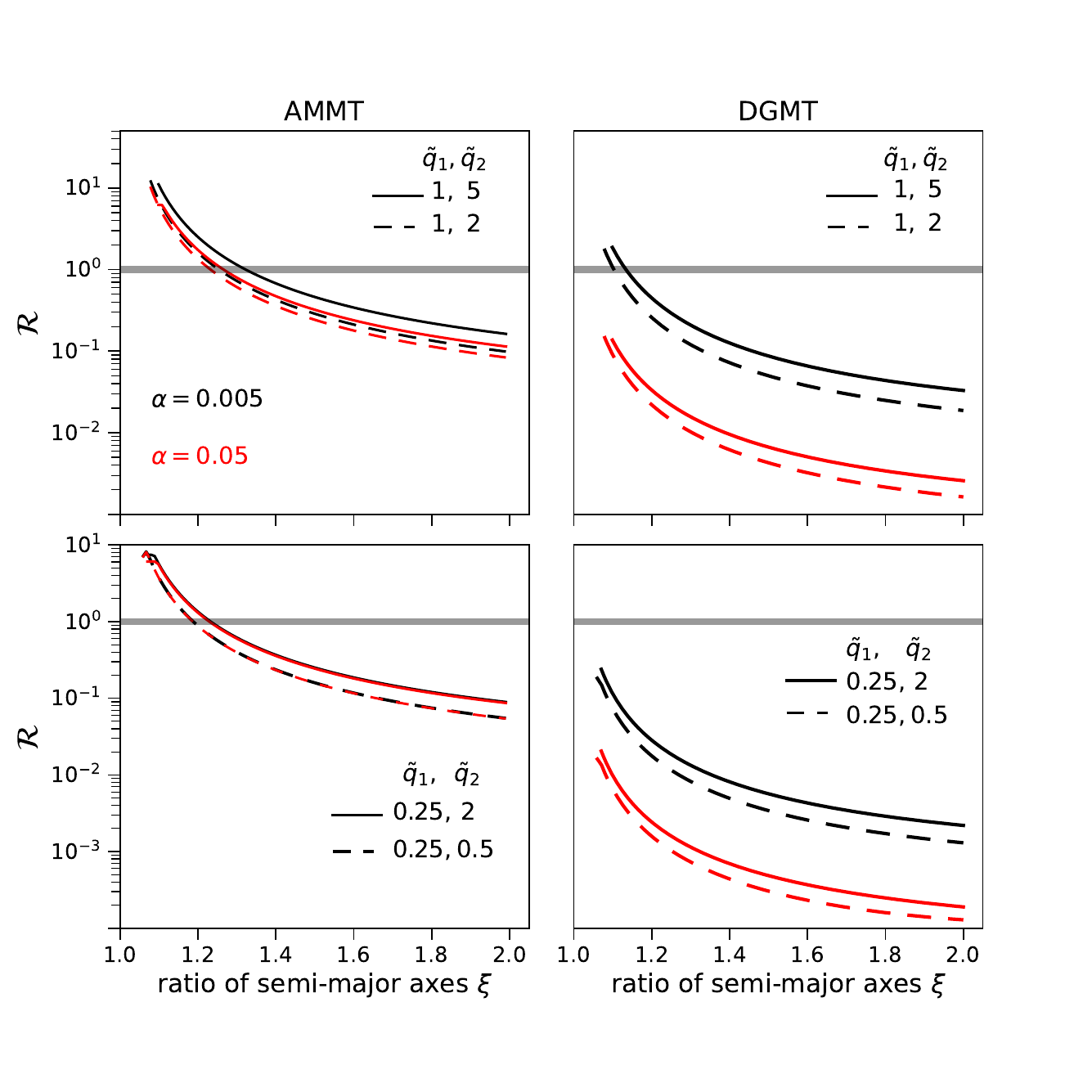}
\vskip -0.3cm
 \caption{Repulsive ratio $\mathcal{R}$ versus the ratio of semi-major axes $\xi$
for different combinations of $\tilde{q}_{1}$,
$\tilde{q}_{2}$ and $\alpha$ parameter.
The horizontal lines indicate the value $\mathcal{R}=1$, above which the repulsion effect counteracts convergent migration.
The left column corresponds to the AMMT scenario and the right-hand column 
is for the DGMT scenario. The disk has $h=0.028$.
}
 \label{fig:Rcurvilinea_B}
\vskip -0.0cm
 \end{figure}

The repulsion mechanism is significant when $\mathcal{D}_{\rm I}\gtrsim |\mathcal{D}_{0}|$.
If we define $\mathcal{R}$ as the ratio $\mathcal{D}_{\rm I}/|\mathcal{D}_{0}|$, and
using the equation for $\mathcal{D}_{0}$ derived in the Appendix \ref{sec:independent},
and $\mathcal{D}_{\rm I}$ from Equation (\ref{eq:Dint_final}), the above condition implies:
\begin{equation}
\mathcal{R}\equiv \left(\frac{f_{0}}{\chi h \xi^{1/2}}\right)
\frac{\mathcal{D}_{\eta}}
{|\eta \gamma_{2}\xi^{1/2-p}-\gamma_{1}|} \gtrsim 1.
\label{eq:cond1}
\end{equation}
Figure \ref{fig:Rcurvilinea_B} shows $\mathcal{R}$ versus $\xi$ for $h=0.028$ and different combinations of $q_{1}$, $q_{2}$ and $\alpha$. 
For the brevity of the notation, we use the quantities $\tilde{q}_{j}\equiv q_{j}/10^{-5}$.
We consider values of $\xi$ that satisfy the following three conditions. First,
the Hill condition (see Appendix \ref{sec:Hill_condition}).
The second condition is that the horseshoe regions are separated, i.e. $a_{2}-a_{1}>x_{\rm hs,1}+x_{\rm hs,2}$,
where $x_{\rm hs,j}$ is the half-width of the horseshoe
region of orbiter $j$. Finally, to justify the two-dimensional approximation, we demand that $a_{2}-a_{1}>2H_{12}$, where $H_{12}$ is the vertical scale height of the disk at $R=a_{12}$, where $a_{12}\equiv (a_{1}+a_{2})/2$.

As expected, $\mathcal{R}$ decreases as the separation between the orbiters increases.
For the parameters under consideration in Figure \ref{fig:Rcurvilinea_B}, 
$\mathcal{R}\lesssim 1$ in DGMT. In AMMT, the condition $\mathcal{R}>1$ 
is fulfilled at $\xi\lesssim 1.3$.

It is interesting to note that, for the case $\tilde{q}_{1}=0.25$ (corresponding to the bottom row of Fig.~\ref{fig:Rcurvilinea_B}),
the repulsive effect is a bit larger for $\tilde{q}_{2}=2$ than for $\tilde{q}_{2}=0.5$, in both AMMT and DGMT. This result implies that although
the inward migration rate of the outer perturber is larger for $\tilde{q}_{2}=2$ than for $\tilde{q}_{2}=0.5$, this increase is correspondingly lower than the increase in the inward migration rate of the inner perturber. 

AMMT predicts that when $\tilde{q}_{1}$ and
$\tilde{q}_{2}$ are sufficiently small, $\mathcal{R}$ is almost independent of $\alpha$
(see lower left panel of Figure \ref{fig:Rcurvilinea_B}). The trend is different in DGMT; $\mathcal{R}$ decreases as $\alpha$ increases, even if $\tilde{q}_{1}$ and $\tilde{q}_{2}$ are small.

In the case that the inner body does not migrate because it is in a migration trap, the corresponding interaction offset is
\begin{equation}
\mathcal{D}_{\rm I}= \lambda_{2}M_{2}^{-1}|T_{1s,1}|,
\end{equation}
and the condition $\mathcal{D}_{\rm I}\gtrsim |\mathcal{D}_{0}|$ is simplified to
\begin{equation}
\mathcal{R}\equiv \frac{f_{0} \lambda_{2} \gamma_{1}\xi^{p-1}}{ \chi h \eta^{2}
\gamma_{2}}\gtrsim 1.
\label{eq:cond1_trap}
\end{equation}
We have checked that if all the parameters ($\tilde{q}_{1}, \tilde{q}_{2}, h, \alpha, \xi$) are fixed, $\mathcal{R}$ is significantly larger when both orbiters can migrate.

\
\section{The simulations}
\label{sec:simulations}
In this section, we compute the disk forces acting on a pair of massive bodies
embedded in a two-dimensional disk,
using the publicly available code FARGO \citep{mas00}. We aim to explore the separations between the pair at which the repulsion effect is important and to test whether AMMT or DGMT can correctly predict the scaling and magnitude of ${\mathcal{D}}_{\rm I}$.
In order to isolate how the disk torques change in a disk already disturbed by its
companion, the bodies are forced to move on fixed Keplerian 
orbits with orbital radii $a_{1}$ and $a_{2}$, and eccentricities $e_{1}$ and $e_{2}$ (constant
over time).

In all our models, the initial (unperturbed) radial profile of the surface density of the disk,
$\Sigma_{\rm un}(R)$, follows the power-law assumed in Section~\ref{sec:predictions}.
For simplicity, $h$ is taken to be constant with $R$ and over time (locally isothermal
disk) and can take one of two values: $h=0.028$ and $h=0.05$. 
We will consider two values for $\alpha$ ($0.005$ and $0.05$) in our simulations.
We limit ourselves to mass ratios $\tilde{q}_{j}\leq 5$. For this range of masses and the values of $h$ under consideration, the horseshoe regions of the orbiters do not overlap if $\xi> 1.09$.

To model the gravitational potential of the orbiters, we introduced a gravitational softening length $R_{s}=0.6H$.
The forces on each perturber are calculated by summing the gravitational force in each grid cell. We use two tapering Gaussian
functions to reduce the contribution of material bound to the orbiters
\begin{equation}
f_{j}(s_{j}) = 1-\exp[-s_{j}^{2}/R_{H,j}^{2}],
\end{equation}
where $\vecs_{j}=\vecr-\vecr_{j}$ and 
$R_{H,j}= (q_{j}/3)^{1/3} r_{j}$ is the individual Hill radius of orbiter $j$.
Accretion onto the orbiters is not included in our simulations.

In code units, the inner boundary is located at $R=0.35$ and the outer boundary at $R=3$, with wave-killing boundary conditions \citep{bor06}. In all the simulations with two perturbers, the initial semi-major axis of the 
inner orbiter is $a_{1}=0.93$. The surface density of the disk is scaled so that the disk mass contained within
$R=0.93$ is $2.4\times 10^{-3} M_{\bullet}$.
In code units, $M_{\bullet}=1$ and the orbital period is $2\pi$ at $R=1$.
We measure the time in terms of the orbits revolved by a body at $R=1$.

\begin{figure}
\hskip -0.1cm
  \includegraphics[angle=0,width=85mm,height=72mm]{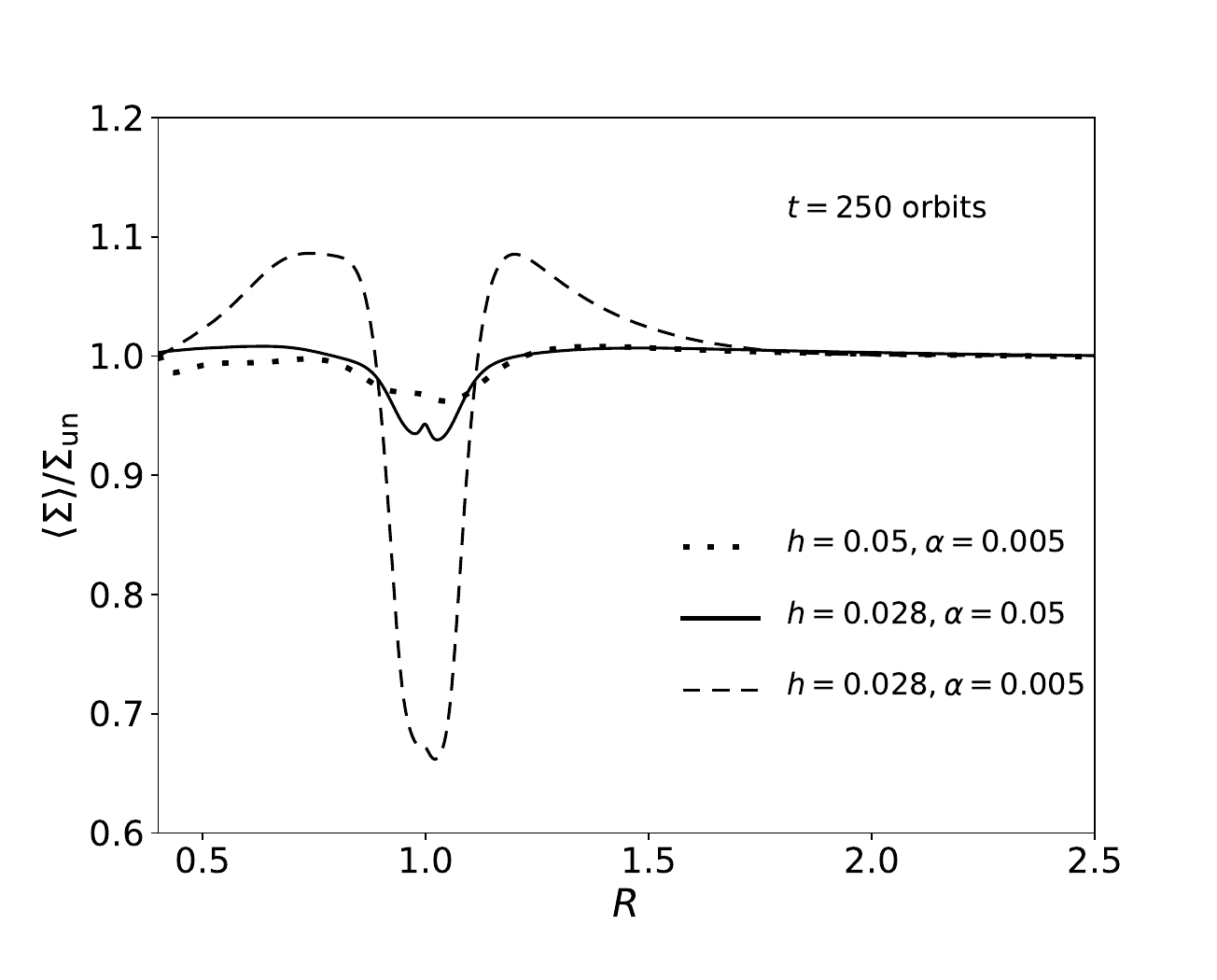}
 \caption{Ratio between the azimuthally-averaged surface density 
$\left<\Sigma\right>(R)$ at $t=250$ orbits and its initial value $\Sigma_{\rm un}(R)$. 
A single object on a fixed circular orbit with $\tilde{q}=5$ was inserted in the disk. Different curves
correspond to different values of $h$ and $\alpha$.
}
 \label{fig:Sigma_1part}
 \end{figure}

In all our simulations, the grid has $N_{R}=950$ (logarithmically
spaced) and $N_{\phi}=2400$ zones in the radial
and azimuthal directions, respectively. 

\begin{figure*}
\hskip -1.2cm
  \includegraphics[angle=0,width=209mm,height=60mm]{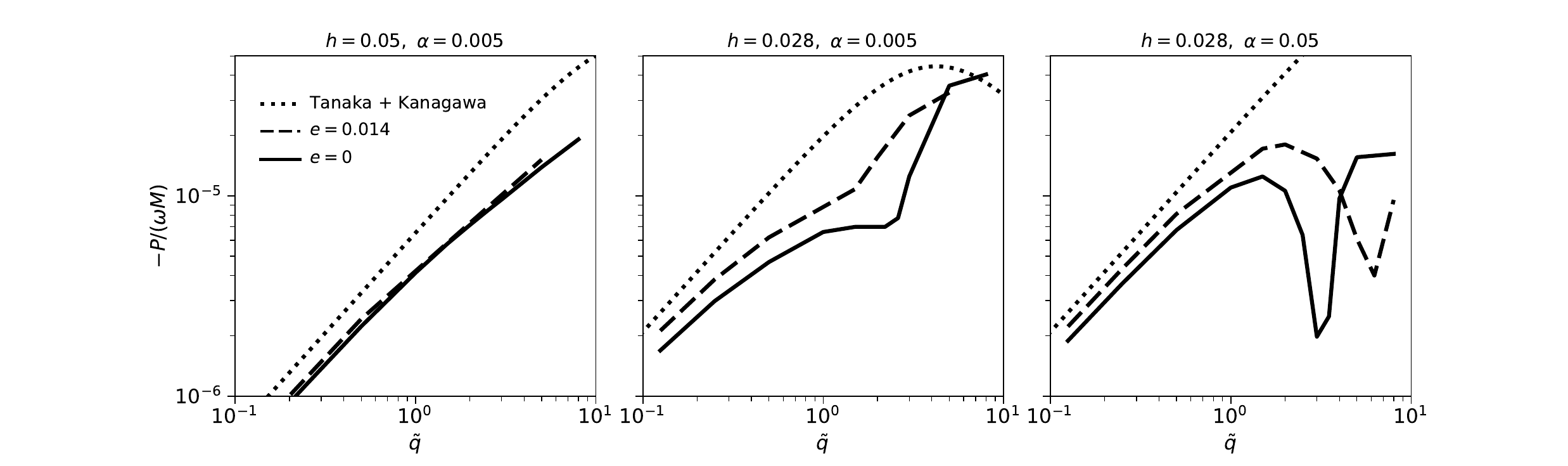}
 \caption{Magnitude of $P/(\omega M)$ when only one perturber with mass ratio
$\tilde{q}$ is inserted
in the disk. The object is in circular orbit (solid lines) or quasi-circular orbit
($e=0.014$; dashed lines) with semi-major axis $a=1$. The dotted lines represent
the formula $-P/(\omega M)= \chi \gamma h^{-2} q \omega^{2} a^{4}\Sigma_{0}M_{\bullet}^{-1}$, which corresponds to the value 
for a perturber in a circular orbit in an inviscid disk \citep{tan02}, including
a factor $\gamma$ as suggested by \citet{kan18} (see  
Appendix \ref{sec:independent}). The values of $h$ and
$\alpha$ of the disk vary from panel to panel.
}
 \label{fig:tq_1part}
 \end{figure*}

A certain model is specified by seven dimensionless parameters $\tilde{q}_{1}$, $\tilde{q}_{2}$, $\xi$, $e_{1}$, $e_{2}$, $h$ and $\alpha$. Throughout this section, however, we use either $\xi$ or $\Delta/H_{12}$, with $\Delta\equiv a_{2}-a_{1}$, and
remind that $H_{12}$ is the vertical scale-height of the disk at $R=a_{12}$.
The relation between $\xi$ and $\Delta/H_{12}$ is given by
\begin{equation}
\frac{\Delta}{H_{12}}=\frac{2(\xi-1)}{(\xi+1) h}. 
\end{equation}

\begin{figure}
\vskip -.1cm
\hskip 0.0cm
\includegraphics[angle=0,width=90mm,height=70mm]{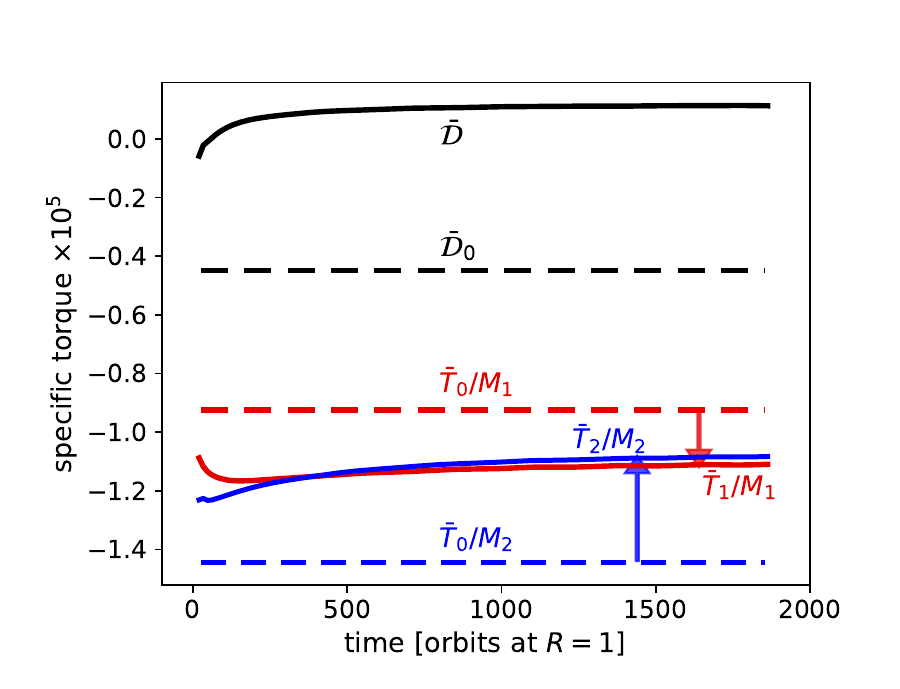}
 \caption{Specific torques acting on the inner orbiter (solid red line), which has $\tilde{q}_{1}=3$, and on the outer perturber (solid blue line), which has $\tilde{q}_{2}=5$. The solid black line represents 
$\bar{\mathcal{D}}\equiv \bar{\mathcal{D}}_{0}+\bar{\mathcal{D}}_{\rm I}$, whereas
the dashed lines labeled with $\bar{T}_{0}/M_{1}$ and $\bar{T}_{0}/M_{2}$ 
indicate the torques when the orbiters are isolated in the disk. 
The arrows indicate the change in the torque due to the presence of the companion.
The disk has $h=0.05$ and $\alpha=0.005$.
The orbits are circular with a separation $\Delta=3H_{12}$ (or, equivalently, $\xi=1.16$). 
}
 \label{fig:paardekooper}
 \end{figure}

\begin{figure*}
\vskip -.3cm
\hskip 0.4cm
\includegraphics[angle=0,width=174mm,height=138mm]{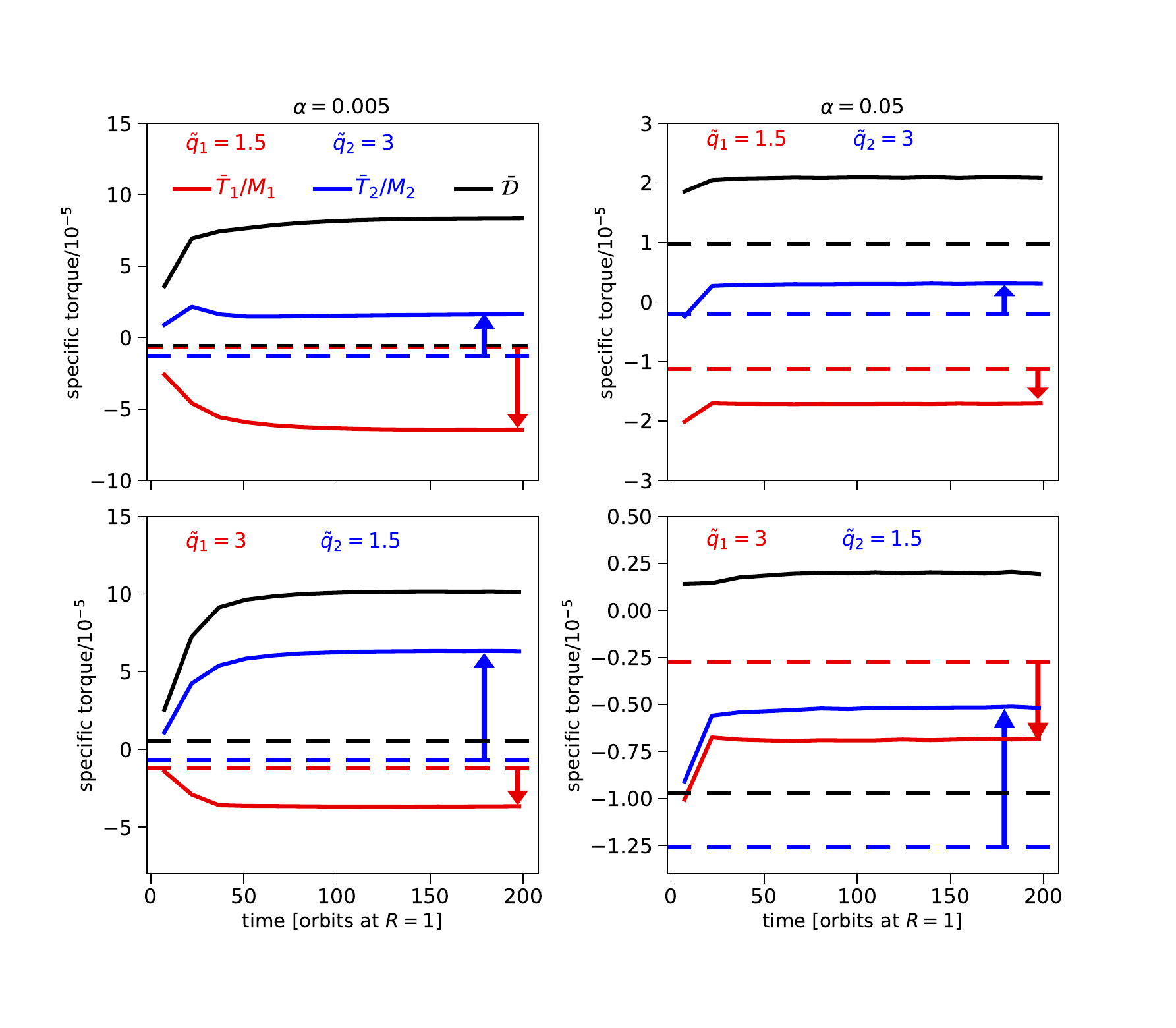}
 \caption{
Similar as Fig. \ref{fig:paardekooper} but for $h=0.028$ and $\Delta=3.1H_{12}$ (which
corresponds to $\xi=1.09$ in this case), and different
combinations of  $q_{1}$ and $q_{2}$. In the left column, we take $\alpha=0.005$,
whereas $\alpha=0.05$ in the right-hand column. 
The dashed lines indicate the values when the orbiters are isolated in the disk. 
Recall that the solid black lines represent $\bar{\mathcal{D}}\equiv \bar{\mathcal{D}}_{0}+\bar{\mathcal{D}}_{\rm I}$.
Note that the blue arrows are always pointing upward (the outer torque is less negative), whereas the 
red arrows are always pointing downwards (the inner torque becomes more negative).
}
 \label{fig:tq_vs_time}
 \end{figure*}

\begin{figure*}
\includegraphics[angle=0,width=198mm,height=55mm]{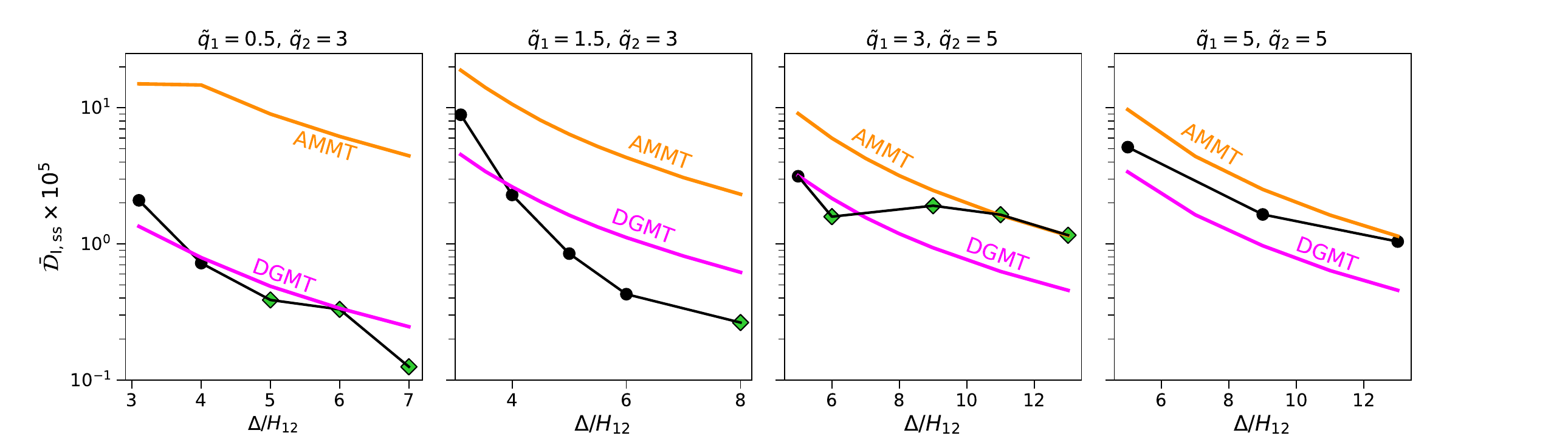}
\hskip -2.8cm
\vskip -0.25cm
 \caption{$\bar{\mathcal{D}}_{\rm I,ss}$ versus $\Delta/H_{12}$ 
as obtained in the simulations (symbols), together with the predicted values in
AMMT (orange lines) and DGMT (magenta lines), in a disk with $h=0.028$ and
$\alpha=0.005$. Black dots indicate divergent migration ($\bar{\mathcal{D}}_{\rm ss}>0$),
whereas green diamonds indicate that the migration is convergent ($\bar{\mathcal{D}}_{\rm ss}<0$).
}
 \label{fig:Dint_vs_xi}
\end{figure*}

\begin{figure*}
\includegraphics[angle=0,width=198mm,height=55mm]{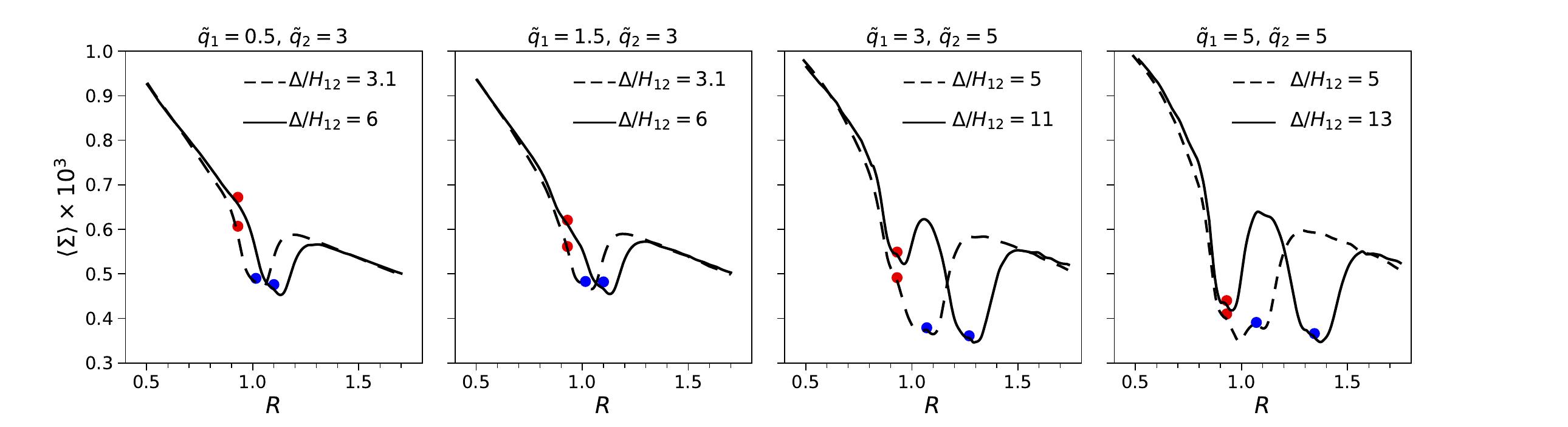}
\hskip -2.8cm
\vskip -0.25cm
 \caption{Azimuthally-averaged surface density $\left<\Sigma\right>$
after $1000$ orbits at $R=1$. The position of the inner and outer orbiters are marked with red and blue dots,
respectively.
In all the cases $h=0.028$ and $\alpha=0.005$.
}
 \label{fig:Sigma_vs_R}
\end{figure*}

\begin{figure*}
\includegraphics[angle=0,width=178mm,height=60mm]{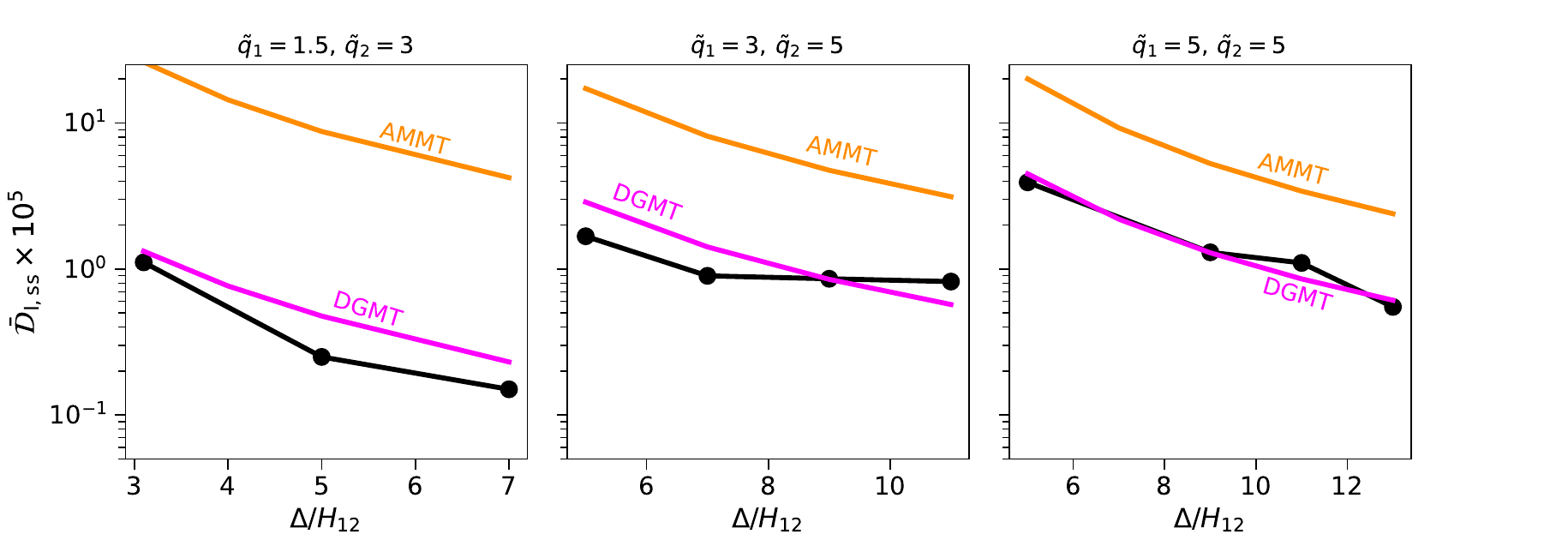}
\hskip 0.8cm
\vskip -0.05cm
 \caption{
Similar to Fig. \ref{fig:Dint_vs_xi} but for $\alpha=0.05$ (again $h=0.028$).
We do not show, however, the corresponding panel for $\tilde{q}_{1}=0.5$ and $\tilde{q}_{2}=3$ because $D_{\rm I,ss}$ becomes very small, almost compatible with zero.
In all the simulations in this figure, it holds $\bar{\mathcal{D}}_{\rm ss}>0$.
}
 \label{fig:Dint_vs_xi_alphaL}
\end{figure*}

\subsection{Individual orbiter}
\label{sec:individual_BH}

We start by considering the case where a single object with mass 
ratio $\tilde{q}=5$ is on a fixed circular orbit with semi-major axis $a=1$.
Figure \ref{fig:Sigma_1part} shows the ratio between the azimuthally-averaged surface density $\left<\Sigma\right>$ at $t=250$ orbits and the initial surface density, for different
combinations of $h$ and $\alpha$. We see that the gap depth is small for $h=0.05$ and $\alpha=0.005$ and $h=0.028$ and $\alpha=0.05$.
However, for $h=0.028$ and $\alpha=0.005$, a partial gap ($\Sigma_{\rm gap}/ \Sigma_{0}\simeq 0.66$) is carved in the disk.  
Since we restrict ourselves to models with $\tilde{q}_{j}\leq 5$, our orbiters 
open shallow or partial gaps when they are alone in the disk.

Figure \ref{fig:tq_1part} shows the steady-state magnitude of the specific power,
more specifically $P/(\omega M)$, versus $\tilde{q}$, for two orbital eccentricities. 
In the cases with $h=0.05$ and $\alpha=0.005$ (left panel), $P/(\omega M)$ is a monotonic function of $\tilde{q}$.
For $h=0.028$ and $\alpha=0.05$, $P/(\omega M)$ presents a gap.
The position of the gap in $P$ depends on the orbital eccentricity;  it is located at 
$\tilde{q}\simeq 3$ for $e=0$ and at $\tilde{q}\simeq 8$ for $e=0.014$.
The origin of this gap in the torque in high-viscosity disks was already discussed in \citet{mak06}.

\subsection{Torques on a pair: Fixed circular orbits}
In this section, we will focus on the particular case where the two components of the pair move on fixed circular orbits ($e_{1}=e_{2}=0$, at any time) around the
central object.

\subsubsection{Temporal behaviour of $\mathcal{D}$}
When two orbiters are immersed in the disk, 
the disk forces onto one of them display large variations in time because
of the interaction with the perturbed density field created by the other one. Thus, we compute the torques using time averages over $N_{\rm av} \tau_{\rm syn }$ where $N_{\rm av}$ is an integer and $\tau_{\rm syn}$ is the time between two successive passages of the orbiters through opposition (or synodic period):
\begin{equation}
\tau_{\rm syn}= \frac{2\pi}{\omega_{1}-\omega_{2}}.
\end{equation}
We will use a bar over a quantity (e.g.~$\bar{T}_{j}$ and $\bar{\mathcal{D}}$) to denote 
its time-average over $N_{\rm av}\tau_{\rm syn}$.

\begin{figure}
\vskip -.34cm
  \includegraphics[angle=0,width=80mm,height=76mm]{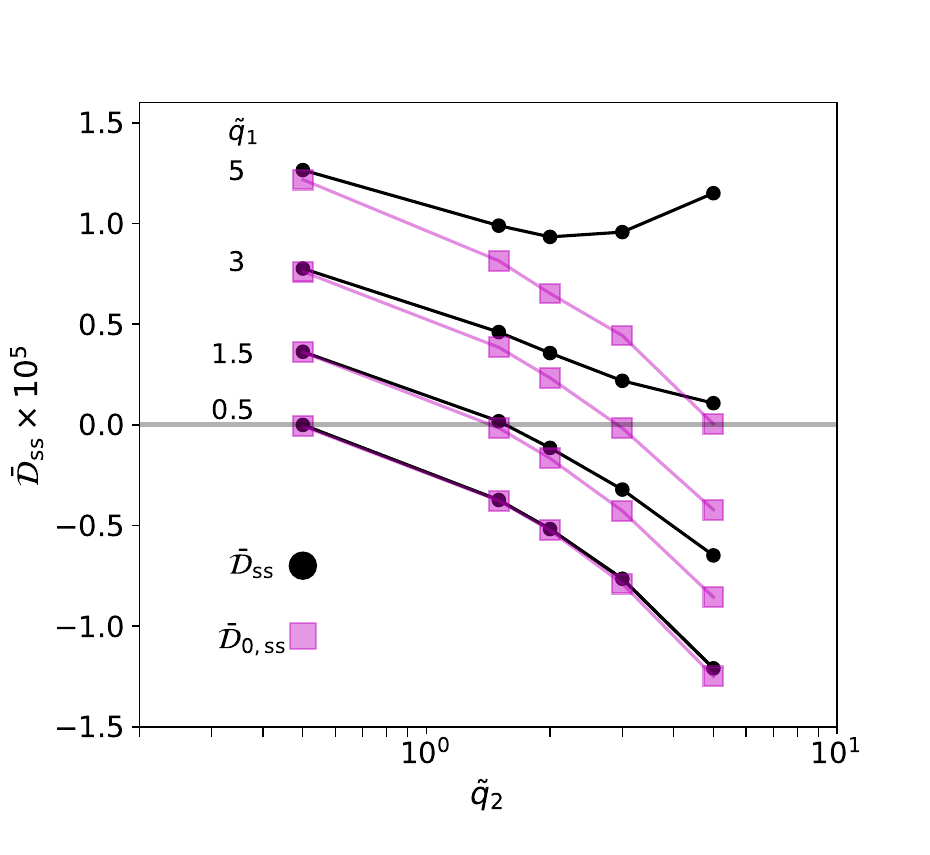}
\hskip -1.0cm
  \caption{$\bar{\mathcal{D}}_{\rm ss}$ (dots) and $\bar{\mathcal{D}}_{\rm 0,ss}$
(squares) along curves of constant $\tilde{q}_{1}$ as a function of
$\tilde{q}_{2}$, in a disk with 
$h=0.05$ and $\alpha=0.005$. We take $\Delta=3H_{12}$, which corresponds to $\xi=1.16$. 
  }
  \label{fig:D_h005_alphas}
\end{figure}

Figures \ref{fig:paardekooper} and \ref{fig:tq_vs_time} show representative examples of the temporal evolution of $\bar{T}_{j}/M_{j}$ and $\bar{\mathcal{D}}$, for $h=0.05$ and $h=0.028$, respectively, in the circular case ($e_{1}=e_{2}=0$). 
After sufficiently long times, $\bar{T}_{j}$ and $\bar{\mathcal{D}}$  reach an almost constant value. We will denote these ``steady-state'' values with the subscript ss, e.g. $\bar{T}_{1,\rm ss}$, $\bar{T}_{2,\rm ss}$ and $\bar{\mathcal{D}}_{\rm ss}$.

For orbiters in circular orbits with separations $\Delta=3.1H_{12}$, or in disks with $\alpha=0.05$, our simulations are long enough to reach a state where $\bar{\mathcal{D}}$ (and also the torques) is almost constant. Therefore, $\bar{\mathcal{D}}_{\rm ss}$ is computed as the value at those steady states. 
For the simulations with $\alpha=0.005$ and $\Delta =5H_{12}$, we ran the simulations to at least $800$ orbits.
If the torques had not reached a steady-state value, we continued the simulations until $\bar{\mathcal{D}}$ varied less than $4\%$ for the last $100$ orbits. 
Then, we took $\bar{\mathcal{D}}_{\rm ss}$ as the value
of $\bar{\mathcal{D}}$ at the end of the simulation.

\begin{figure}
\hskip -0.42cm
\includegraphics[angle=0,width=93mm,height=155mm]{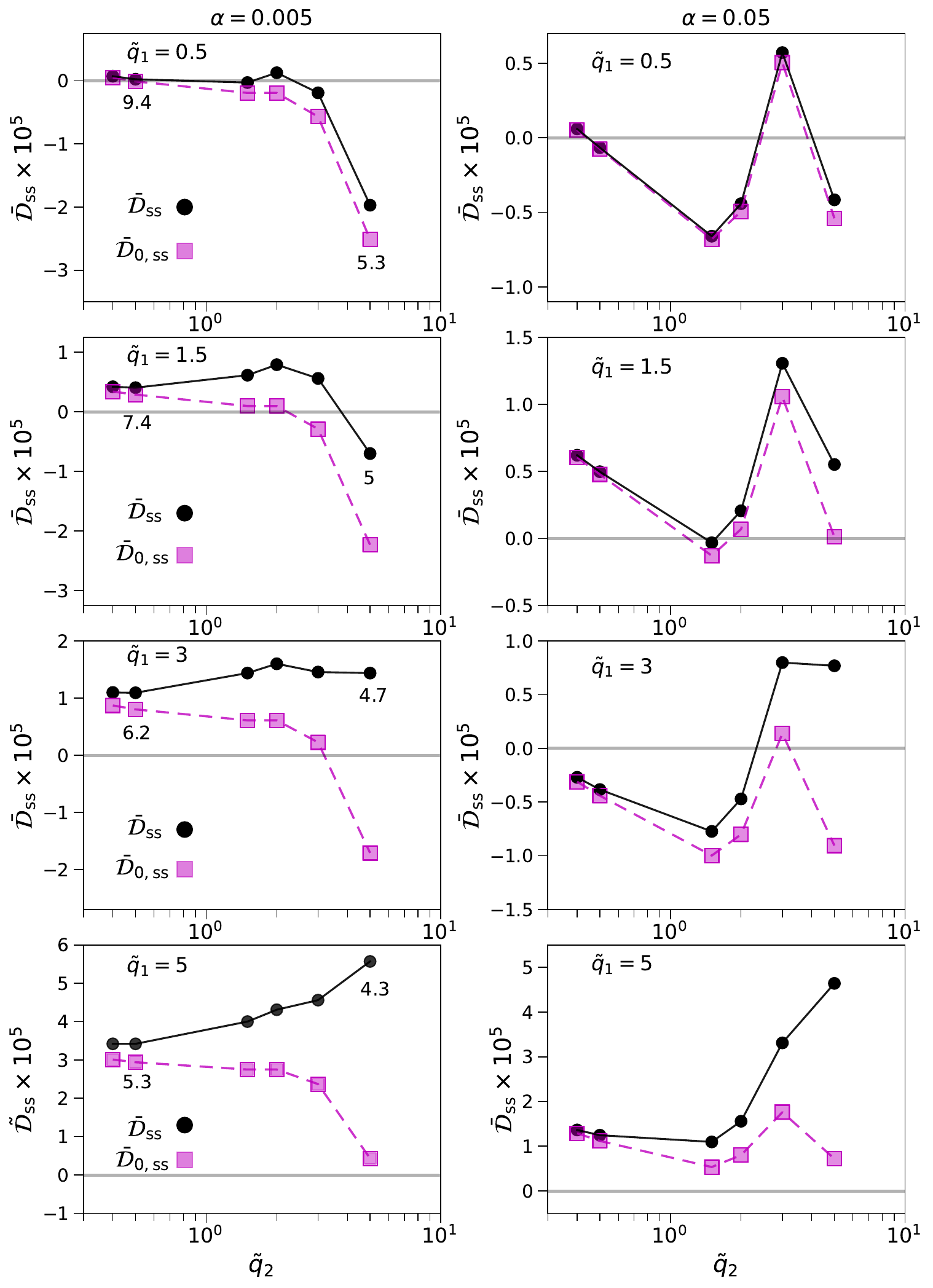}
\vskip -0.0cm
 \caption{$\bar{\mathcal{D}}_{\rm ss}$ (dots) and $\bar{\mathcal{D}}_{\rm 0,ss}$
(squares), as in Figure \ref{fig:D_h005_alphas}, but now in a disk with $h=0.028$,
and two different values of $\alpha$: $0.005$ (left column) and $0.05$ (right-hand column).
The orbital separation is $\Delta=5H_{12}$ (or, equivalently, $\xi=1.15$).  
The numbers below the dots at $\tilde{q}_{2}=0.5$ and at $\tilde{q}_{2}=5$ in the left panels indicate the mutual Hill separation 
$\equiv (a_{2}-a_{1})/R_{\rm mH}$; all the experiments satisfy the Hill stability condition.
}
  \label{fig:D_h0028_2alpha_5H}
\end{figure}
\begin{figure}
\hskip -0.65cm
\includegraphics[angle=0,width=93mm,height=150mm]{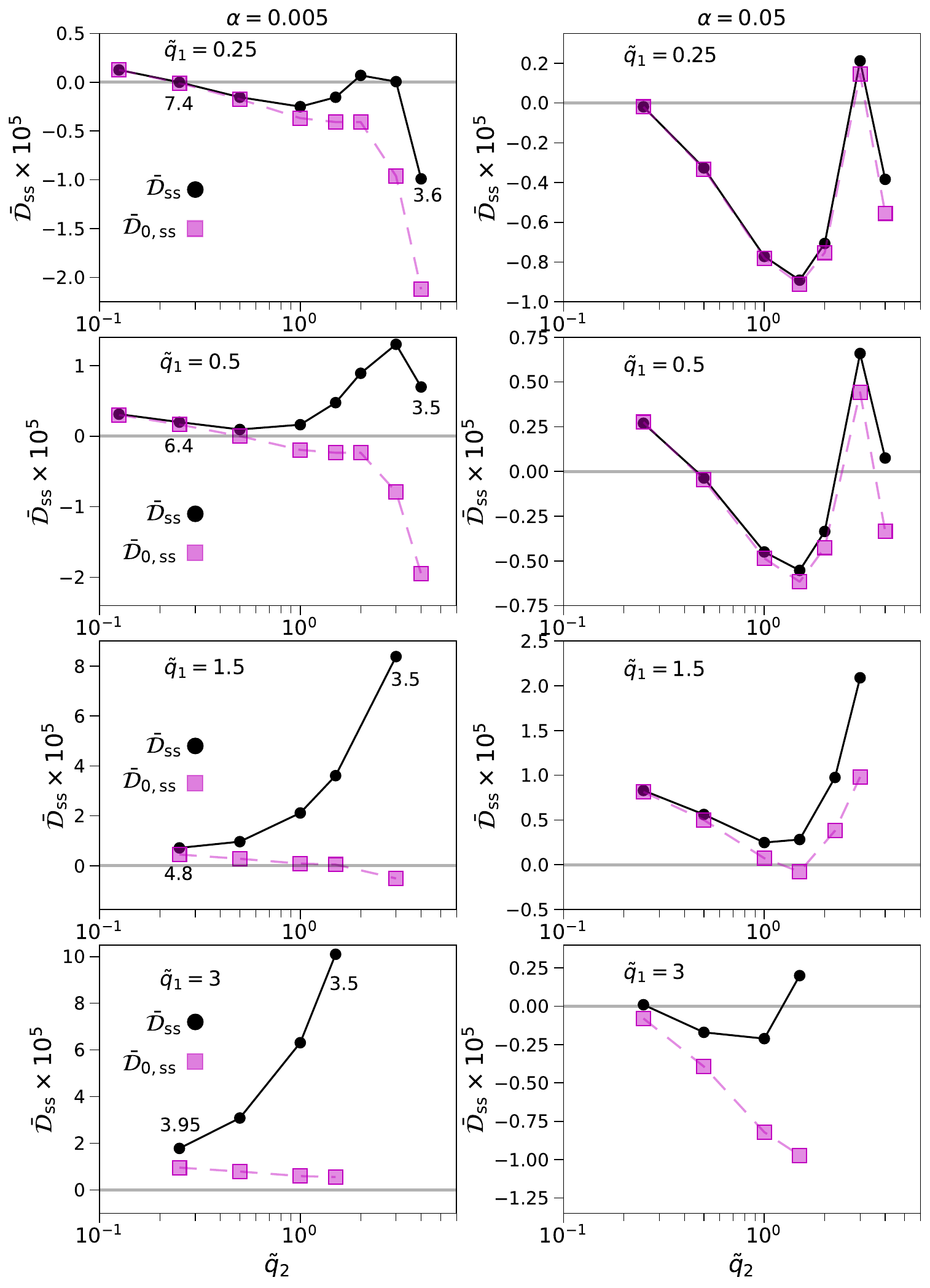}
\vskip -0cm
 \caption{$\bar{\mathcal{D}}_{\rm ss}$ (dots) and $\bar{\mathcal{D}}_{\rm 0,ss}$
(squares)
for different combinations of $\tilde{q}_{1}$ and $\tilde{q}_{2}$, separated by 
$\Delta=3.1H_{12}$ in a disk with 
$h=0.028$ (i.e. $\xi=1.09$). The viscosity parameter $\alpha$ is $0.005$ in the left panels and 
$0.05$ in the right panels.  
For reference, we give the mutual Hill separation for the first and last points on the left
panels; all the experiments
satisfy the Hill stability condition. 
 }
  \label{fig:D_h0028_2alpha_3H}
\end{figure}

\
\subsubsection{$\bar{\mathcal{D}}_{\rm I,ss}$ versus $\Delta$}
\label{sec:int_vs_Delta}
It is worthwhile to look again at Figures \ref{fig:paardekooper} and \ref{fig:tq_vs_time} 
to gain insight into the repulsion effect. In all cases, $\bar{T}_{1,\rm ss}$ is more negative (the magnitude of the torque is larger) than it is when the orbiter is isolated (red horizontal lines).
On the other hand, $\bar{T}_{2,\rm ss}$ is always more positive than the torque when it is treated as a single body (blue horizontal lines). Therefore, the disk-mediated 
interaction between orbiters leads to a repulsive effect. Interestingly, in all cases presented in Figures \ref{fig:paardekooper} and \ref{fig:tq_vs_time}, $\bar{\mathcal{D}}_{\rm ss}>0$.

The interaction offset in the steady state, $\bar{\mathcal{D}}_{\rm I,ss}$, can be computed in our simulations as  $\bar{\mathcal{D}}_{\rm I,ss} = \bar{\mathcal{D}}_{\rm ss}-\bar{\mathcal{D}}_{\rm 0,ss}$, with $\bar{\mathcal{D}}_{\rm 0,ss}$
calculated from Equation (\ref{eq:def_D}) in simulations where each orbiter is alone in the disk (as those computed in Section \ref{sec:individual_BH}). 

Figure \ref{fig:Dint_vs_xi} shows the dependence of $\bar{\mathcal{D}}_{\rm I,ss}$ on the 
orbital separation $\Delta/H_{12}$ for different combinations of $\tilde{q}_{1}$ and
$\tilde{q}_{2}$. We have taken $h=0.028$, $\alpha=0.005$ and $e_{1}=e_{2}=0$. In all cases shown in Figure \ref{fig:Dint_vs_xi}, the maximum of $\bar{\mathcal{D}}_{\rm I,ss}$ occurs at the smallest value of $\Delta$ considered.
For these lowest values of $\Delta$, $\bar{\mathcal{D}}_{\rm ss}>0$, implying that the migration would be divergent. In general, there is a trend for $\bar{\mathcal{D}}_{\rm I,ss}$ to decrease with the separation (see the left and middle-upper panels). 
However, it becomes almost constant between
$\Delta=6H_{12}$ and $\Delta=13H_{12}$ in the third panel (for $\tilde{q}_{1}=3$ and $\tilde{q}_{2}=5$).

In Figure \ref{fig:Sigma_vs_R}, we present the radial profile of the azimuthally-averaged surface density after $1000$ orbits at $R=1$, for two representative values of the orbital separation $\Delta$. We see that the orbiters share a common
gap for $\Delta\lesssim 6H_{12}$. For the largest orbital separations explored, it is possible to distinguish the ``individual'' gaps (solid lines in the third and fourth panels).
Note that the gap depths in the first and second panels are similar.

In Figure \ref{fig:Dint_vs_xi}, we have also shown $\bar{\mathcal{D}}_{\rm I,ss}$ predicted in AMMT and DGMT described in Section \ref{sec:theory} and Appendices \ref{sec:app_AMMT} and \ref{sec:app_DGMT}.
We see that AMMT overestimates $\bar{\mathcal{D}}_{\rm I,ss}$ up to one order of magnitude in some cases. 
DGMT, on the other hand, can predict the value of $\bar{\mathcal{D}}_{\rm I,ss}$ within a factor of $3$. It can also reproduce the slight ascend (from left to right-hand panels) of the curves from the simulations.

Figure \ref{fig:Dint_vs_xi_alphaL} shows $\bar{D}_{\rm I,ss}$ versus $\Delta$, as in Figure \ref{fig:Dint_vs_xi}, but now for $\alpha=0.05$. 
We see that $\bar{\mathcal{D}}_{\rm I,ss}$ for $\tilde{q}_{1}=1.5$ and $\tilde{q}_{2}=3$ is generally smaller for $\alpha=0.05$ than for $\alpha=0.005$. However, in the two models with $\tilde{q}_{1}\geq 3$ and $\tilde{q}_{2}=5$, $\bar{\mathcal{D}}_{\rm I,ss}$ for $\alpha=0.05$ is quite similar to that for $\alpha=0.005$.

In the cases shown in Figure \ref{fig:Dint_vs_xi_alphaL}, AMMT overestimates $\bar{\mathcal{D}}_{\rm I,ss}$
and predicts that $\bar{\mathcal{D}}_{\rm I,ss}$ should be larger as viscosity increases, but the simulations do not show that trend. On the other hand, DGMT reasonably predicts the magnitude of the repulsive effect.
A more comprehensive comparison between the results of simulations and the models will be presented in Section
\ref{sec:comparison}.

\
\subsubsection{Dependence of the repulsion effect on $\tilde{q}_{1}$ and $\tilde{q}_{2}$}

To see whether the orbiters can reverse the direction of migration from converging
to diverging, it is worthwhile to compare $\bar{\mathcal{D}}_{\rm ss}$ with $\bar{\mathcal{D}}_{\rm 0,ss}$. Figure \ref{fig:D_h005_alphas} shows $\bar{\mathcal{D}}_{\rm 0,ss}$ and  $\bar{\mathcal{D}}_{\rm ss}$, both obtained from the simulations, for $h=0.05$ and $\alpha=0.005$, when $\Delta=3H_{12}$ (implying $\xi=1.16$). Along the curves, $\tilde{q}_{2}$ is varied, keeping $\tilde{q}_{1}$ constant.  
We see that for $\tilde{q}_{1}=0.5-1.5$ and for the values of $\tilde{q}_{2}$ under consideration,  $\bar{\mathcal{D}}_{\rm ss}$ is rather similar to $\bar{\mathcal{D}}_{\rm 0,ss}$, meaning that the repulsion is small. The curves for $\bar{\mathcal{D}}_{\rm ss}$ and for $\bar{\mathcal{D}}_{\rm 0,ss}$ essentially overlap for $\tilde{q}_{2}=0.5$ and separate from each other as $q_{2}$ increases. 
We find that for $\tilde{q}_{1}\geq 3$ and $\tilde{q}_{2}\geq 3$, the disk-mediated interaction between orbiters should be taken into account, for separations $\xi = 1.16$. 
Yet for these values of $\tilde{q}_{1}$ and $\tilde{q}_{2}$, the depth of the common gap is still very shallow.
For instance, for $\tilde{q}_{1}=\tilde{q}_{2}=5$, the depth of the gap is only  $\Sigma_{\rm gap}/\Sigma_{\rm un, gap}=0.935$.

Figure \ref{fig:D_h0028_2alpha_5H} shows $\bar{\mathcal{D}}_{\rm ss}$ and  $\bar{\mathcal{D}}_{\rm 0,ss}$ for $h=0.028$ and $\Delta=5H_{12}$. In this case, the semi-major axis ratio $\xi=1.15$ is very similar to the value adopted in Figure \ref{fig:D_h005_alphas}.
We find that $\bar{\mathcal{D}}_{\rm ss}>\bar{\mathcal{D}}_{\rm 0,ss}$, implying that
the effect is repulsive. 
$\bar{\mathcal{D}}_{\rm I,ss}$ increases as $\tilde{q}_{2}$ increases (fixed $\tilde{q}_{1}$) and as $\tilde{q}_{1}$ increases (fixed $\tilde{q}_{2}$). In some cases, $\bar{\mathcal{D}}_{\rm 0,ss}<0$, but $\bar{\mathcal{D}}_{\rm ss}>0$.

Consider first the low viscosity disk ($\alpha=0.005$; left panels in Fig. \ref{fig:D_h0028_2alpha_5H}). 
For $\tilde{q}_{2}=2$, $\bar{\mathcal{D}}_{\rm ss}>0$ for any value of $\tilde{q}_{1}\in [0.5,5]$.
In addition, for $\tilde{q}_{1}\geq 1.5$ and $\tilde{q}_{2}\leq 3$, we have $\bar{\mathcal{D}}_{\rm ss}>0$.
In particular, if we take $\tilde{q}_{1}=1.5$ and $\tilde{q}_{2}=2.4$, similar to those in Kepler-36 for illustration, we find that the migration is divergent if $\Delta=5H_{12}$, provided that $h=0.028$, $\alpha=0.005$ and the orbits are circular.

In a disk with $h=0.028$ and $\alpha=0.05$ (right panels in Fig. \ref{fig:D_h0028_2alpha_5H}), the curves for $\bar{D}_{\rm 0,ss}$ exhibit a peak at $\tilde{q}_{2}=3$; this is reminiscent of the reduction of the torque for this value of the mass ratio (see Figure \ref{fig:tq_1part}). $\bar{D}_{\rm ss}$ also shows this peak but only when $\tilde{q}_{1}\leq 1.5$. 
For $\tilde{q}_{1}\leq 1.5$ and $\tilde{q}_{2}\leq 3$,  $\bar{D}_{\rm ss}\simeq \bar{D}_{\rm 0,ss}$. Interestingly, for $\tilde{q}_{1}= 1.5-3$ and $\tilde{q}_{2}=5$, we have that $\bar{D}_{\rm 0,ss}<0$ (convergent migration), whereas $\bar{D}_{\rm ss}>0$ (divergent migration).

Figure \ref{fig:D_h0028_2alpha_3H} shows the results for $\Delta=3.1H_{12}$, again for $h=0.028$ (for these parameters, $\xi=1.09$). Quite remarkably, even for the smallest inner mass considered in this figure, the repulsive effect is sufficiently strong to yield a divergent migration for $\tilde q_2=2$, in the low viscosity case ($\bar{D}_{\rm ss}>0$). The migration is also divergent in all the models explored in that figure  with $\alpha=0.005$ and $\tilde{q}_{1}\geq 0.5$. For  $\alpha=0.005$, $\tilde{q}_{1}=0.5$ and $\tilde{q}_{2}\gtrsim 0.5$, migration switches from convergent to divergent because $\bar{\mathcal{D}}_{\rm 0,ss}<0$ and $\bar{\mathcal{D}}_{\rm ss}>0$. 

Finally, we focus on models with $\alpha=0.05$ (right panels in Figure \ref{fig:D_h0028_2alpha_3H}). For $\tilde{q}_{1}\leq 1.5$, the shape of the curves for $\bar{D}_{\rm ss}$ is rather similar to the shape of $\bar{D}_{\rm 0,ss}$ curves. However, for $\tilde{q}_{1}=3$, the large discrepancy between $\bar{D}_{\rm 0,ss}$ and $\bar{D}_{\rm ss}$ indicates that the disk-mediated interaction plays an important role.

\begin{figure}
\hskip -0.0cm
\includegraphics[angle=0,width=90mm,height=146mm]{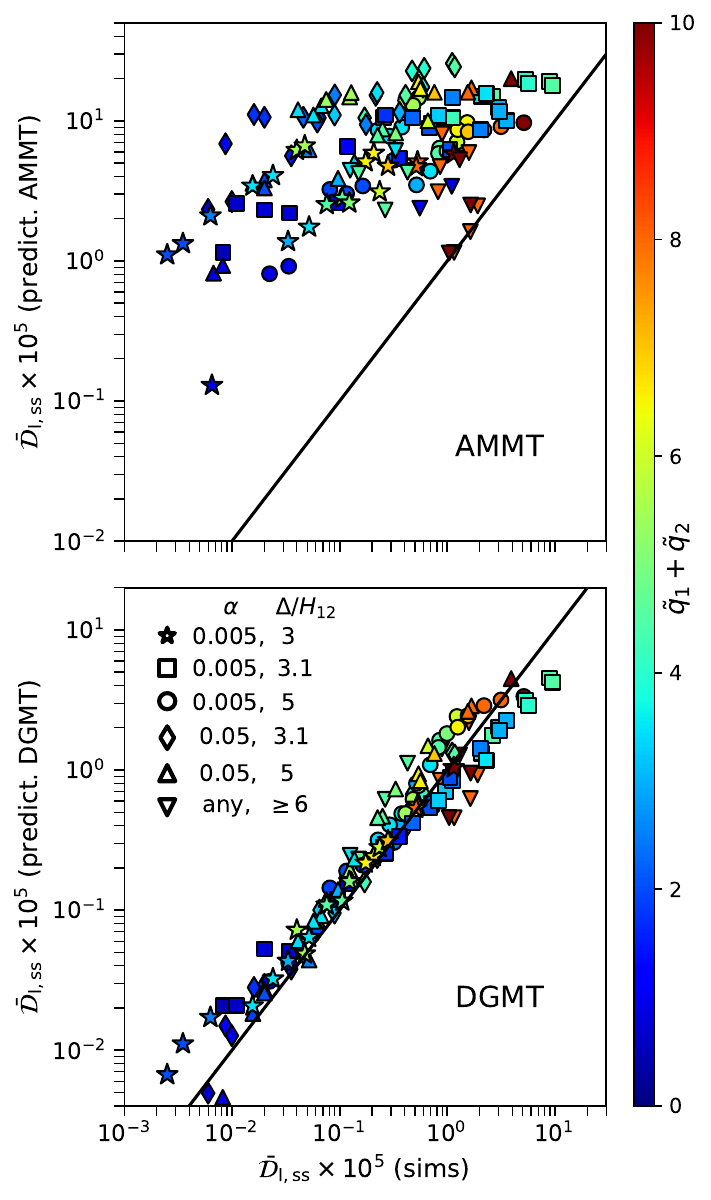}
 \caption{The interaction offset $\bar{\mathcal{D}}_{\rm I,ss}$ in the numerical simulations shown in Figs. \ref{fig:D_h005_alphas}, \ref{fig:D_h0028_2alpha_5H} and \ref{fig:D_h0028_2alpha_3H}, is compared with the results of AMMT (upper panel) and DGMT (lower panel). 
The attributes of the symbols are given in the legend in the lower panel.
The color bar indicates $\tilde{q}_{1}+\tilde{q}_{2}$.
The solid lines represent the identity function.}
 \label{fig:comparison_theor_sims}
 \end{figure}

\begin{figure}
\hskip -0.0cm
  \includegraphics[angle=0,width=90mm,height=72mm]{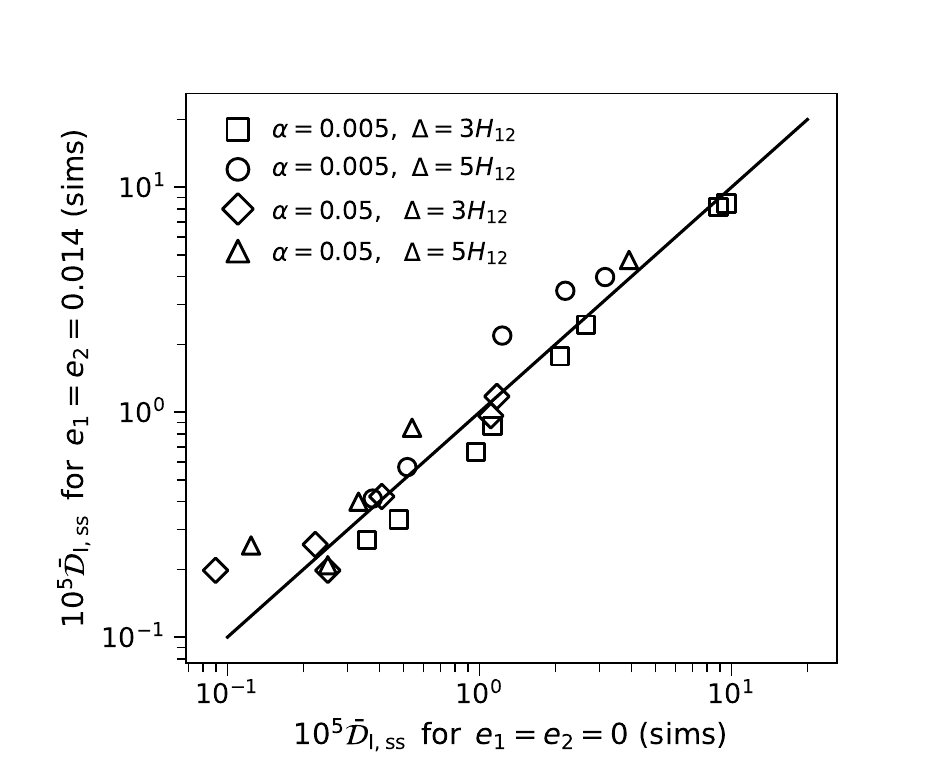}
 \caption{A comparison of $\bar{\mathcal{D}}_{\rm I,ss}$ in simulations with $e_{1}=e_{2}=0.014$ and in simulations with $e_{1}=e_{2}=0$.  In all the experiments $h=0.028$.
The solid line represents the identity function.}
 \label{fig:ecc0014}
 \end{figure}

\begin{figure}
  \includegraphics[angle=0,width=82mm,height=70mm]{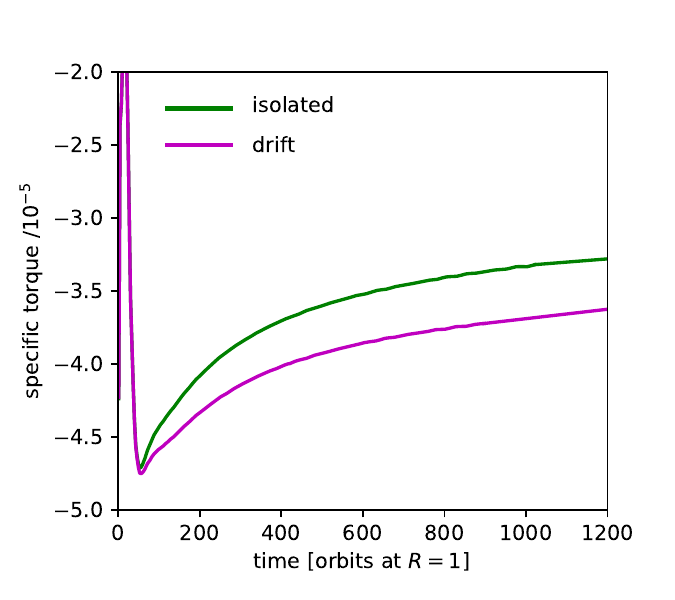}
\hskip 1.0cm
\vskip -0.0cm
  \caption{Specific torque exerted by the disk on an orbiter in two different cases: (1) without any companion (green curve), and (2) when the orbiter is alone in the disk but an
external positive torque $\Lambda_{\rm imp}$ is applied to the disk (magenta curve).}
 \label{fig:drift}
\end{figure}

\subsubsection{Comparison with the predictions of AMMT and DGMT}
\label{sec:comparison}
In Section \ref{sec:int_vs_Delta}, we already mentioned that, for the cases considered in that section, DGMT is more consistent with the results of simulations than AMMT, as AMMT generally overestimates the magnitude of $\bar{\mathcal{D}}_{\rm I,ss}$.
This is confirmed in Figure \ref{fig:comparison_theor_sims} where we show a one-to-one comparison between the predicted values of $\bar{\mathcal{D}}_{\rm I,ss}$ and those obtained from the simulations. 
We see that AMMT fails to predict the scalings of 
$\bar{\mathcal{D}}_{\rm I,ss}$ and, in addition, overestimates $\bar{\mathcal{D}}_{\rm I,ss}$ for orbital separations $\leq 5H_{12}$. However,
DGMT correctly captures the scaling of $\bar{\mathcal{D}}_{\rm I,ss}$ with $h$, $\alpha$, $\tilde{q}_{1}$, $\tilde{q}_{2}$ and radial separation $\Delta$. When $\bar{\mathcal{D}}_{\rm I,ss}\lesssim 0.4\times 10^{-5}$, 
the simulations show a mildly weaker effect than predicted by DGMT. This has a simple explanation. 
The time scale to reach a steady state is $\sim 200$ orbits for $\alpha=0.05$, and $\sim 2000$ orbits
for $\alpha\sim 0.005$
\citep[e.g.,][]{ata18}. For $\alpha=0.005$, the time is a bit longer than the running times of our simulations. 
This is particularly true for low-mass perturbers ($\tilde{q}_{1}+\tilde{q}_{2}< 2$) in a disk with $h=0.05$
and $\alpha=0.005$ (the three star symbols at the bottom left corner of Figure \ref{fig:comparison_theor_sims}),
for which our simulations were not long enough to establish a steady state. As a consequence, the variations of the slope of surface density have not reached their steady state values, hence numerical simulations show a repulsion weaker than predicted.
On the other hand, the scatter in the upper part of the diagram (at values $\bar{\mathcal{D}}_{\rm I,ss}\gtrsim 0.4\times 10^{-5}$) may reflect the fact that the analytical model for the gap profile given by Equation (\ref{eq:delta_Sigma1}), as well as the expression for
the corotation torque in Equation (\ref{eq:CR_torque}) used to compute $\bar{\mathcal{D}}_{\rm I,ss}$ in DGMT become less accurate for masses larger than the thermal mass ($q\gtrsim h^{3}$).

\subsection{Eccentric orbits}
\label{sec:eccentric}
In the previous section, we assumed that the orbiters have null eccentricities. However, the gravitational interaction between the pair's components will excite their eccentricities. Therefore, it is worthwhile to see how sensitive $\bar{\mathcal{D}}_{\rm I,ss}$ is to the orbiters' eccentricities.
To do so, we have conducted the following test. For a subset of the simulations shown in Figs. \ref{fig:D_h0028_2alpha_5H} and \ref{fig:D_h0028_2alpha_3H}, we have redone the simulations with the same parameters but adopting $e_{1}=e_{2}=h/2=0.014$ for the orbital eccentricities of the perturbers, which were kept constant throughout the duration of the runs. We ran the models for the same number of orbits as we did in their circular counterparts and computed $\bar{\mathcal{D}}_{\rm ss}$ as the average of $\bar{\mathcal{D}}$ over the last $3$ complete oscillations (typically $\simeq 100$ orbits). Since the shape of the radial profile of the gaps open by individual orbiters with $e=0.014$ is rather similar to that for individual orbiters in a circular orbit, we expect that DGMT will be for these eccentric pairs as accurate as for circular pairs. 
Figure \ref{fig:ecc0014} compares  
$\tilde{\mathcal{D}}_{\rm I,ss}$ as obtained in the simulations with $e_{1}=e_{2}=0$ with
the values in simulations
having $e_{1}=e_{2}=0.014$. We see that $\tilde{\mathcal{D}}_{\rm I, ss}$ may 
differ by up to $50\%$ for the models under consideration.
In summary, as long as the orbital eccentricities are moderate ($e\leq h/2$), 
$\bar{\mathcal{D}}_{\rm I,ss}$ hardly changes by a factor larger than $2$ as compared to the case where the eccentricities are fixed to zero.

\section{Discussion: why does AMMT yield wrong results?} 
\label{sec:discussion}
In Section \ref{sec:comparison} we have found that AMMT generally overestimates 
the effect of repulsion, implying that the orbiters absorb less angular momentum than the amount deposited in their horseshoe regions. Here, we present a simple simulation that clearly shows that the excess torque
on the orbiter is not that deposited in the disk within its horseshoe region. More specifically, we conducted a simulation 
of the outer orbiter alone in the disk but applying an imposed (external) positive torque
density, $\Lambda_{\rm imp}(R)$, to the disk. This external torque imitates the torque deposited
by the inner orbiter.
To avoid spurious changes in the surface density due to the imposed torque, 
we require that the specific torque $\Lambda_{\rm imp}/\Sigma$ scales with $R$ as $\propto B(R)/\Sigma_{\rm un}(R)$, where $B(R)=\Omega(R)/4$ is the second Oort constant,
with $\Omega(R)$ the disk orbital frequency \citep{mas03}. In the absence of the orbiter,
this torque induces a constant outward drift in a disk with a stationary axisymmetric
density profile $\Sigma_{\rm un}(R)$. In this particular choice of $\Lambda_{\rm imp}(R)$,
no waves are excited by the external torque.

We have computed the torque on an orbiter with $\tilde{q}_{2}=5$ at $a_{2}=1.34$,
in a simulation where it is isolated in the disk (with $h=0.028$ and $\alpha=0.005$), 
and in a simulation where
the effect of the inner orbiter has been replaced by 
$\Lambda_{\rm imp}(R)$, with $\Lambda_{\rm imp}=0.8\times 10^{-8}$ (in code units)
at $R=a_{2}$. 
This magnitude of the external torque density corresponds to the torque density deposited
at $R=1.34$ by the damping of the wake excited by an inner orbiter with $\tilde{q}_{1}=5$
at $a_{1}=0.93$, when alone in the disk. 
It was measured in the simulations of a single orbiter, as $\Lambda_{\rm dep,1}(R)=-d\dot{J}_{\rm w}/dR$, where  $\dot{J}_{\rm w}$ is the 
angular momentum flux carried by the waves, which is given by
\begin{equation}
\dot{J}_{\rm w}(R)=R^{2} \int_{0}^{2\pi} (v_{\phi}-\bar{v}_{\phi}) (v_{R}-\bar{v}_{R}) \Sigma \,d\phi,
\end{equation}
where $v_{R}$ and $v_{\phi}$ are the radial and azimuthal components of the gas velocity, respectively, and the bar indicates the azimuthal average.
We comment that this procedure neglects a potential excitation term (by the inner perturber) at the orbit of the outer perturber. The ratio of orbital radii between the outer and inner perturber is $1.44$ and the ratio of orbital periods is $1.72$, which indicates that all the outer Lindblad resonances of the inner perturber are located inside of the outer perturber's orbit, except that of the $m=1$ Fourier component of the inner planets' potential, which lies outside. The residual excitation that may take place over the width of the horseshoe region should therefore represent a tiny fraction of incident flux of angular momentum, especially in a disc as thin as the one considered in this experiment. Furthermore, these considerations about the magnitude of the deposition of angular momentum within the horseshoe region will soon appear futile, as the torque excess on the outer orbiter does not even match the sign of the torque deposited.

Figure \ref{fig:drift} shows the torques as a function of time. According to AMMT, the magnitude of the torque acting on the outer orbiter should be augmented by an amount of $2\Lambda_{\rm dep,1}R_{{\rm hs}, 2}$ (evaluated at $R=a_{2}$) with respect to the isolated case. This implies an offset in the specific torque $T_{2}/M_{2}$ of $\sim 2\times 10^{-5}$ if $\tilde{q}_{2}=5$. This result is in sharp contrast with the outcomes of the simulation with $\Lambda_{\rm imp}$, where we see that the torque presents an offset of $\sim -0.3\times 10^{-5}$ after $800$ orbits (the torque is more negative than it is when the orbiter is isolated in the disk; see magenta and green curves in Fig. \ref{fig:drift}). 
However, this result is in agreement with studies of type~III migration, where the relative planet-disk drift is achieved by inducing a disk drift under the action of external torque. The torque applied to the disk entails an outward drift. This is equivalent to a situation where the disk does not drift, and the planet migrates inwards. If the horseshoe region is partially depleted, as is the case here, type~III torques arise and exert a positive feedback on migration \citep{mas03}, i.e. they tend to make the negative torque larger in absolute value. This is precisely what we see in Fig.~\ref{fig:drift}. This goes against the expectation that the torque on the orbiter due to the action of the external torque can be evaluated by calculating the angular momentum given to the disk within its horseshoe region. 
While it is true that fluid elements trapped in the horseshoe region ultimately transmit the positive torque deposited to the orbiter, fluid elements flowing (outwards) through the horseshoe region extract angular momentum from the orbiter during their unique horseshoe U-turn with the latter. The net effect is an increase in the absolute value of the torque.

The standard AMMT scenario contemplates only the contribution from within the horseshoe region and arbitrarily discards the effect of material immediately interior to that region that flows towards the outer disk. By construction, this scenario yields a torque excess that has the correct sign, but it cannot have the correct value, as it considers only half of the problem.

\section{Summary}
\label{sec:summary}

In this work, we have re-examined the repulsion effect in packed pairs in cases where the pair components carve partial gaps in the disk ($\Sigma_{\rm gap}/\Sigma_{\rm un,gap}\gtrsim 0.5$). We focus on cases where the total mass of the pair is $\lesssim 10^{-4}M_{\bullet}$. In the case of protoplanetary disks, this corresponds to sub-Neptune-mass planets. In the case of AGN disks, this corresponds to the so-called extreme mass ratio inspirals (EMRIs). It includes an inspiral formed by $(10-300) M_{\odot}$ BHs in an accretion disk around a central $10^{6-7} M_{\odot}$ massive BH. 

We have used $\bar{\mathcal{D}}_{\rm I,ss}$ as a measure of the repulsion effect.
We have developed a semi-analytical framework to provide a quantitative view of the
scalings of $\bar{\mathcal{D}}_{\rm I,ss}$ with the orbiters' mass, orbital separation, disk aspect ratio, and viscosity. We applied two different assumptions: AMMT and DGMT. AMMT assumes that the torques are modified because each orbiter absorbs the angular momentum deposited on its horseshoe region by the wakes excited by its respective companion. DGMT assumes that the corotation torques are modified because the companion modifies the surface density profile of the disk.
The formulation can also be applied when the inner orbiter does not migrate if it is in a migration trap.

We have measured $\bar{\mathcal{D}}_{\rm I,ss}$ in two-dimensional simulations of a pair of orbiters in fixed orbits for a wide range of orbiters' mass ratios ($0.1\lesssim q_{2}/q_{1}\lesssim 10$), and disk viscosity (a variation of a factor of $10$).
The magnitude of $\bar{\mathcal{D}}_{\rm I,ss}$ spans nearly three orders of magnitude. We have confirmed, in agreement with previous results, that (1) high-mass orbiters present a larger repulsion effect than low-mass orbiters, (2) the repulsive effect may be large enough to stall convergent migration, regardless of whether the pair is close or not to resonance, and (3) the repulsion effect decreases as the disk viscosity is increased.
In most of our simulations, we forced the perturbers to be on circular orbits. Nevertheless, we have also explored orbits with eccentricities of $0.5h$ and found similar repulsion.

We have compared the predictions of AMMT and DGMT with the results of the hydrodynamical simulations. AMMT is incapable of reliably predicting the scaling of $\bar{\mathcal{D}}_{\rm I,ss}$ with $q_{1}$, $q_{2}$, $h$ and $\alpha$. In particular, for our adopted range of parameters, AMMT generally overestimates the magnitude of the wake-orbiter interaction by up to two orders of magnitude, especially in disks with high viscosity. 

We find that a more robust approach is DGMT, which can successfully account for 
the scaling of $\bar{\mathcal{D}}_{\rm I,ss}$, at least for the range of parameters explored in this paper. Note that this work focuses exclusively on the repulsion effect due to the forces
mediated by the disk, ignoring the mutual gravitational interaction between the pair
components, which can be strong, especially when the planets are near first-order commensurabilities.

\begin{acknowledgements}
We would like to thank the referee for a thorough reading of the manuscript and constructive comments that improved the quality of the paper.

{\it Software:} FARGO \citep{mas00}.
\end{acknowledgements}

\begin{appendix}

\section{$\lambda_{1}$ and $\lambda_{2}$ in the AMMT assumption}
\label{sec:app_AMMT}
In this Appendix, we derive the angular momentum deposited in the horseshoe regions.
The angular momentum flux excited by orbiter $j$ can be written as 
$\Phi_{j}(R) T_{{\rm 1s},j}$, with $\Phi_{j}(R)$ the dimensionless angular momentum flux.
For isolated perturbers that do not open deep gaps, 
$\Phi(R)$ can be found in \citet{raf02} \citep[see also][]{goo01}.
It can be approximated by:
\begin{equation}
\Phi_{j}=
\left\{
\begin{array}{cl}
1 &\textup{if  }  \,\,\tau_{j}(R)<\tau_{{\rm sh},j} \\
  &  \\

\sqrt{\tau_{\rm sh}/\tau_{j}} &\textup{if  }  \,\,  \tau_{j}(R)>\tau_{{\rm sh},j},
\end{array}
\right.
\label{eq:damping_Rafikovcc}
\end{equation}
where $\tau_{{\rm sh},j}$
is the dimensionless wave-to-shock timescale (see \citet{raf02}
for details), given by 
\begin{equation}
\tau_{{\rm sh},j}=1.89+0.53 h^{3}/q_{j},
\label{eq:shock_position}
\end{equation}
and
\begin{equation}
\tau_{j}(R) = \frac{3}{(2 h^{2})^{5/4}} \bigg|\int_{1}^{R/a_{j}} |s^{3/2}-1|^{3/2}
s^{(p-3)/2} ds \bigg|.
\label{eq:rafikov43cc}
\end{equation}
We recall that $p$ is the power-law exponent of the unperturbed surface density of the disk.
The fraction of the angular momentum excited by the perturber $3-j$ that is deposited in the horseshoe region of perturber $j$ is
\begin{equation}
\lambda_{j}'=|\Phi_{3-j}(b_{j}^{+})-\Phi_{3-j}(b_{j}^{-})|
\label{eq:net}
\end{equation}
where $b_{j}^{\pm}=a_{j}\pm x_{{\rm hs},j}$,  with 
$x_{{\rm hs},j}\simeq a_{j}\sqrt{q_{j}/h}$ the
half-width of the horseshoe region.

In principle, the deposition torque depends on the surface density profile of the disk. Equations (\ref{eq:damping_Rafikovcc})-(\ref{eq:net}) assume that the radial profile of the surface density follows a power-law and ignore the fact that the companion could open a gap that modifies the underlying structure of the disk. For perturbers that open partial gaps, this non-linear effect is likely small \citep[e.g.,][]{gin18}.

Following \citet{cui21}, we will assume that the perturber can absorb the angular momentum deposited in its horseshoe region if it is less than the one-sided horseshoe drag $T_{\rm hs}$. Using that
$|T_{{\rm hs},j}|= 0.5\gamma_{j}\Sigma_{{\rm un},j} \omega^{2} a_{j}^{4}(q_{j}/h)^{3/2} = 
0.5 h^{-3/2}\gamma_{j} \omega_{0}^{2}a_{0}^{3+p}
a_{j}^{1-p} \Sigma_{0} q_{j}^{3/2}$, the condition $|\lambda_{3-j}T_{\rm{1s}, j}|
\leq |T_{{\rm hs}, 3-j}|$ implies:
\begin{equation}
\lambda_{1}={\rm min}\left[\lambda_{1}', \,\frac{h}{f_{0}} \frac{\gamma_{1}}{\gamma_{2}}
\left(\frac{h}{q_{1}}\right)^{1/2} \eta^{-2} \xi^{p-1}\right],
\end{equation}
and
\begin{equation}
\lambda_{2}={\rm min}\left[\lambda_{2}', \,\frac{h}{f_{0}} \frac{\gamma_{2}}{\gamma_{1}}
\left(\frac{h}{q_{2}}\right)^{1/2} \eta^{2} \xi^{1-p}\right].
\end{equation}

\section{$\lambda_{1}$ and $\lambda_{2}$ in the DGMT assumption}
\label{sec:app_DGMT}

In this Appendix, we compute the change in the corotation torque as the companion modifies the disk surface density.

The unsaturated corotation torque on perturber $1$ in the steady state case, when it is alone in the disk on a fixed circular orbit, is given by 
\begin{equation}
T_{{\rm CR},1}^{\rm (uns)} = \frac{3}{4} \Sigma_{{\rm un}, 1} \omega_{1}^{2} x_{{\rm hs},1}^{4} 
\frac{d\ln (\Sigma_{\rm un}/B)}{d\ln R}\bigg|_{R=a_{1}},
\label{eq:CR_torque}
\end{equation}
where $B$ is the second Oort's constant \citep{gol79,war91,war92}.
This formula is valid if the perturber does not open a deep gap in the disk.
In order to take into account that the corotation torque may differ from the
unsaturated value because it can be partially saturated,
we include a correction factor $C$, so that
$T_{{\rm CR},1}=C \cdot T_{{\rm CR},1}^{\rm (uns)}$. 
The correction factor $C$ depends on $z_\nu\equiv a\nu/(\omega x_{\rm hs}^3)$, the ratio between the libration
timescale and diffusion timescale across the horseshoe region. For $z_{\nu}\leq 1$, we take $C$ from \citet{mas10}:
\begin{equation}
    C=\frac{8\pi}{3}z_\nu F(z_\nu),
\end{equation}
where $F$ is defined by Eq.~(120) of \citet{mas10}. For $z_{\nu}>1$, we take the values reported in the 
Figure 12 of the same paper.

If, in the presence of the companion, the surface density profile changes from $\Sigma_{\rm un}(R)$ to $\Sigma_{\rm un}(R)+\delta \Sigma(R)$, the change in the corotation torque will be
\begin{eqnarray}
\Delta T_{{\rm CR},1} &&= \frac{3}{4} C \omega_{1}^{2} x_{{\rm hs},1}^{4} 
\left[ a_{1} \frac{d(\Sigma_{\rm un}+\delta \Sigma)}{dR}\bigg|_{R=a_{1}}+\frac{3}{2} \delta \Sigma_{1} + p \Sigma_{{\rm un},1}\right]\nonumber\\
&& =\frac{3}{4} C\omega_{1}^{2} x_{{\rm hs},1}^{4} 
\left[ a_{1}\frac{d(\delta \Sigma)}{dR} +\frac{3}{2} \delta \Sigma \right] \bigg|_{R=a_{1}},
\label{eq:DeltaTCR}
\end{eqnarray}
where we have assumed that the rotation of the disk is Keplerian (i.e. $d\ln B/d\ln R=-3/2$).  From \citet{duf15}, we know that 
\begin{equation}
\delta \Sigma (R)= -\Sigma_{\rm un}(R)\left[ \frac{f_{0}\Phi_{2} K_{2}/(3\pi)}{1+f_{0} K_{2}/(3\pi)} \sqrt{\frac{a_{2}}{R}}\right] ,
\label{eq:delta_Sigma1}
\end{equation}
with $K_{2}=q_{2}^{2}/(\alpha h^{5} )$ and $\Phi_{2}(R)$ as given in Equation
(\ref{eq:damping_Rafikovcc}).
Using our definition of $\lambda_{1}\equiv -\Delta T_{{\rm CR},1}/|T_{{\rm 1s,2}}|$ and
recalling that $|T_{{\rm 1s,2}}|=f_{0}\gamma_{2} h^{-3} q_{2}^{2}a_{2}^{4}\omega_{2}^{2}
\Sigma_{{\rm un}, 2} $, we obtain
\begin{equation}
\lambda_{1} = -\frac{3}{4} \frac{C h^{3}}{f_{0}} \frac{\xi^{p}}{\gamma_{2}q_{2}^{2}}  
\frac{x_{{\rm hs},1}^{4}}{a_{1}^{3}a_{2}}
\left[ \frac{a_{1}}{\Sigma_{{\rm un},1}}\frac{d(\delta \Sigma)}{dR} +\frac{3}{2} 
\frac{\delta \Sigma}{\Sigma_{{\rm un},1}} \right] \bigg|_{R=a_{1}}.
\end{equation}

A similar procedure can be followed to derive
$\lambda_{2}\equiv \Delta T_{{\rm CR},2}/|T_{{\rm 1s},1}|$. We get
\begin{equation}
\lambda_{2} = \frac{3}{4} \frac{Ch^{3}}{f_{0}} \frac{\xi^{-p}}{\gamma_{1}q_{1}^{2}} 
\frac{x_{{\rm hs},2}^{4}}{a_{1}a_{2}^{3}}
\left[ \frac{a_{2}}{\Sigma_{{\rm un},2}}\frac{d(\delta \Sigma)}{dR} +\frac{3}{2} 
\frac{\delta \Sigma}{\Sigma_{{\rm un},2}} \right] \bigg|_{R=a_{2}}.
\end{equation}
Here $\delta \Sigma$ is given by Equation (\ref{eq:delta_Sigma1}) but replacing the 
subscript $2$ with $1$. In the cases of interest, $\lambda_{1}$ and $\lambda_{2}$
are both positive, so the effect is generally repulsive. As expected, $\lambda_{1}$
and $\lambda_{2}$ are proportional to the fourth power of $x_{{\rm hs},1}$ and $x_{{\rm hs},2}$, respectively.
Thus, to predict correctly the magnitude of $\lambda_{1}$ and
$\lambda_{2}$, it is important to have a precise value for $x_{{\rm hs},j}$.
The procedure to compute $x_{\rm hs}$ in our simulations
is described in the Appendix \ref{app:rhs_sims}.

In principle, the differential Lindblad torque on each orbiter will also be modified if
the structure of the disk changes due to the presence of the companion. We note, however,
that the dependence of the Lindblad torque on the slope of surface density is much weaker than that of the corotation torque, not only in a realistic, 3D disk
\citep{tan02}, but also in the 2D disks considered in this paper, even for the low-mass planets for which the width of the horseshoe region is not enhanced (see Appendix \ref{app:rhs_sims}). 
In fact, \citet{paa10} showed that the differential Lindblad torque, $T_{\rm L}$,
on perturber $j$ (alone in the disk) in circular orbit in a 2D disk with constant $h$ 
is given by
\begin{equation}
\gamma \frac{T_{{\rm L},j}}{T_{0,j}}=\left(-4.2-0.1\frac{d\ln\Sigma_{\rm un}}{d\ln R}\bigg|_{R=a_{j}}\right)\left(\frac{0.4}{R_{s}/H}\right)^{0.71},
\end{equation}
with $T_{0,j}= q_{j}^{2} \Sigma_{{\rm un}, j} a_{j}^{4} \omega_{j}^{2}/h^{2}$. 
From the above equation and for the smoothing length $R_{s}$ used in this work, 
the change in
the differential Lindblad torque when the background density $\Sigma_{\rm un}$ is modified
to $\Sigma_{\rm un}+\delta \Sigma$ is
\begin{equation}
\Delta T_{{\rm L},j}= T_{0,j} \left(-0.075 \frac{a_{j}}{\Sigma_{{\rm un}, j}}
\frac{d(\delta \Sigma)}{dR} -3.1\frac{\delta\Sigma}{\Sigma_{{\rm un},j}}\right)\bigg|_{R=a_{j}}.
\label{eq:DeltaTL}
\end{equation}
We may compare $\Delta T_{{\rm CR}, 1}$ given in Equation (\ref{eq:DeltaTCR}) with $\Delta T_{{\rm L},1}$,
assuming, for simplicity, that the outer perturber opens a Gaussian gap so that 
$\delta \Sigma = -\beta \Sigma_{\rm un} \exp(-(R-a_{2})^{2}/W^{2})$, where $\beta$ is
the depth of the gap and $W$ its width. By combining Eqs. (\ref{eq:DeltaTCR}) and
(\ref{eq:DeltaTL}), it is simple to show that 
\begin{equation}
\frac{|\Delta T_{{\rm L},1}|}{|\Delta T_{{\rm CR},1}|} \simeq \frac{3}{C}\left(1+\frac{2(\xi-1)a_{1}^{2}}{W^{2}}\right)^{-1}.
\end{equation}
To derive this equation, we have taken conservatively that 
$x_{\rm hs,1}\simeq 1.1\sqrt{q_{1}/h}$ in Equation (\ref{eq:DeltaTCR}).
For $W\simeq 3ha_{2}$ and the values of $\xi$ considered in this paper, we find
that $|\Delta T_{{\rm L},1}|\lesssim 0.15 |\Delta T_{{\rm CR},1}|$. A similar argument
leads to $|\Delta T_{{\rm L},2}|\lesssim 0.15 |\Delta T_{{\rm CR},2}|$. It is therefore legitimate to neglect the Lindblad torque in the DGMT scenario.

\section{Two independent migrators}
\label{sec:independent}
In this section, we estimate $\mathcal{D}_{0}$ defined as $\mathcal{D}$, given in Equation (\ref{eq:def_D}), when the bodies are so far apart that the torques they experience are not affected by the presence of their companion. The torque exerted on an isolated migrator in a steady state, i.e.~when the gap has become a steady-state structure, was computed by \citet{kan18}, using two-dimensional hydrodynamical simulations. For $\alpha\geq 5\times 10^{-3}$, they find that the torque
exerted on a migrator that opens a partial gap 
is approximately given by the Type I migration formula but replacing the disk surface density by the surface density at the bottom of the gap 
$\Sigma_{{\rm gap},j}$ (see Figure 7 in Kanagawa et al. 2018\footnote{Note that the points are scattered around the values predicted by this relationship. As discussed by \citet{kan18}, the torque can even be positive in some cases. For the combinations of parameters chosen in this paper, this relationship overestimates $|T_{j}|$ (see Section \ref{sec:individual_BH}).}). From \citet{duf15}, we have $\Sigma_{{\rm gap},j}=\gamma_{j}\Sigma_{{\rm un},j}$, where $\gamma_{j}$ is given in Equation (\ref{eq:gamma_j}). 
On the other hand, we may use the formula for the Type-I torque from \citet{tan02} to obtain  
$T_{j}\simeq -\chi \gamma_{j} h^{-2}q_{j}^{2}  \omega_{j}^{2} a_{j}^{4} \Sigma_{{\rm un},j}$, with $\chi=1.16+2.83p$ in a two-dimensional disk.
Combining these equations, we find from Equation (\ref{eq:def_D}):
\begin{equation}
\mathcal{D}_{0}=\chi h^{-2}\xi^{1/2} \omega_{0}^{2}a_{0}^{3+p} a_{1}^{1-p} \Sigma_{0} M_{\bullet}^{-1}\left(\gamma_{1}q_{1}-\gamma_{2}q_{2}\xi^{1/2-p}\right).
\label{eq:DI_analytic}
\end{equation}
Therefore, if $h$ and $\alpha$ are constant along the disk and
$\gamma_{2}q_{2}/(\gamma_{1}q_{1})=\xi^{p-1/2}$ then 
$\mathcal{D}_{0}=0$ and, consequently, 
$\xi$ remains constant over time (see Eq. \ref{eq:dlnxi_dt}). 
On the other hand, the migration is divergent
if $\gamma_{2}q_{2}/(\gamma_{1}q_{1})<\xi^{p-1/2}$, while it is convergent if 
$\gamma_{2}q_{2}/(\gamma_{1}q_{1})>\xi^{p-1/2}$.

\section{Hill stability condition}
\label{sec:Hill_condition}
Assume that initially, the orbiters are on circular orbits and focus on pairs satisfying the condition for Hill stability. In the absence of the accretion disk, the pair is said to be Hill stable if 
the gravitational interactions between the two orbiters remain moderate. The Hill stability condition is commonly written in terms of the mutual Hill radius, defined as:
\begin{equation}
R_{\rm mH}= \left(\frac{q_{t}}{3}\right)^{1/3} a_{12},
\end{equation}
where $q_{t}\equiv q_{1}+q_{2}$ and $a_{12}\equiv (a_{1}+a_{2})/2$.
In the case of initially circular orbits, \cite{gla93} showed that the condition
$\Delta >2\sqrt{3}R_{\rm mH}$, where $\Delta\equiv a_{2}-a_{1}$, ensures Hill stability because close encounters between the orbiters are forbidden. Even though the interaction between
the orbiters may lead to episodes where their eccentricities can increase, conservation of angular momentum (if the disk is not present) implies that 
the orbital separation increases, impeding close
encounters.
In terms of $\xi\equiv a_{2}/a_{1}$, Gladman's condition implies
\begin{equation}
\xi > \frac{1+3^{1/6}q_{t}^{1/3}}{1-3^{1/6}q_{t}^{1/3}}.
\label{eq:Hill_xi}
\end{equation}
For instance, for $q_{t}=10^{-4}$, it implies $\xi >1.12$.

\section{The half-width of the horseshoe region in our simulations}
\label{app:rhs_sims}
Since some planets in our samples have a mass near or slightly above the thermal mass $h^3M_\bullet$, the standard, low-mass estimate for the width of the horseshoe region may lead to an underestimation of the variation of the corotation torque. It is, therefore, desirable to use a formula for the width of the horseshoe region that captures the transition from the low- to the high-mass regime. \cite{jim17} have studied this transition, but their results apply to planets in three-dimensional disks. Here, we have planets in two-dimensional disks and a smoothing length of the potential of $0.6H$, for which no result exists in the literature. We have therefore undertaken a dedicated study similar to that of \citet{jim17}, but in 2D disks with the smoothing length quoted above. We performed 20 runs over 10 orbital periods, with planet masses ranging from $10^{-6}M_\bullet$ to $10^{-4}M_\bullet$ in a geometric sequence, in an inviscid disk with aspect ratio $h=0.028$. The resolution adopted in these runs was $N_\phi=1000$ and $N_R=400$, the radial bins having a constant spacing covering the range $0.5$ to $1.5$ (the planet's orbital radius being one). The half-width $x_{\rm hs}$ of the horseshoe region is then obtained as the mean of the rear and front upstream half-widths, measured at $\pm 60^\circ$ from the azimuth of the planet. Following \citet{jim17}, expressing the planet's mass $Q$ in units of the thermal mass ($Q=q/h^3$) and the half-width $X_{\rm hs}$ of the horseshoe region in units of the pressure lengthscale $H$ ($X_{\rm hs}=x_{\rm hs}/H$), we obtain the following regimes:
\begin{eqnarray}
    X_{\rm hs}^{\text{low}} &=& 1.11Q^{1/2} \mbox{~~in the low mass limit},\\
   X_{\rm hs}^{\text{high}}&=& 1.6Q^{1/3}\mbox{~~in the high mass limit}.\\
\end{eqnarray}
At any given mass, we find that the horseshoe half width is within $2$~\%, at most, of the linear combination of these two extreme regimes given by:
\begin{equation}
    X_{\rm hs} = \varepsilon X_{\rm hs}^{\mathrm{low}}+(1-\varepsilon) X_{\rm hs}^{\mathrm{high}},
\end{equation}
where $\varepsilon=1/(1+0.7Q^3)$. From this we infer the value $x_{\rm hs}=HX_{\rm hs}$ used in Appendices~\ref{sec:app_AMMT} and~\ref{sec:app_DGMT}.

\end{appendix}

\end{document}